\documentclass[prd,twocolumn,superscriptaddress,floatfix,nopacs,preprintnumbers,nofootinbib]{revtex4}
\usepackage[utf8]{inputenc}

\usepackage{graphicx}
\usepackage[normalem]{ulem}
\usepackage{xcolor}
\usepackage{adjustbox}
\usepackage{mathrsfs}
\usepackage{amssymb,bm}
\usepackage{amsmath}
\usepackage{mathtools}
\usepackage{physics}
\usepackage{slashed}
\usepackage{verbatim}
\usepackage{ragged2e}
\usepackage{caption}
\usepackage{subcaption}
\usepackage{multirow}
\usepackage{array}
\usepackage{tikz}
\usepackage{lipsum} 
\usepackage{soul}
\newcommand{\nc}{N_\mathrm{c}}
\newcommand{\aem}{\alpha_\mathrm{em}}
\newcommand{\gev}{\mathrm{GeV}}

\newcommand{\bt}{\mathbf{b}}
\newcommand{\qso}{Q_{\mathrm{s},0}} 
\newcommand{\qs}{Q_{\mathrm{s}}}

\newcommand{\xt}{\mathbf{x}}

\newcommand{\nf}{n_\mathrm{f}}

\newcommand{\cf}{C_\mathrm{F}}

\newcommand{\xij}[1]{\mathbf{x}_{#1}}

\newcommand{\btheta}{\boldsymbol{\theta}}

\newcommand{\sigmaltnlo}{\sigma_{L,T}^{\textrm{NLO}}}
\newcommand{\sigmaltdip}{\sigma_{L,T}^{\textrm{dip}}}

\newcommand{\sigmaltqgu}{\sigma_{L,T}^{qg, \textrm{unsub.}}}
\newcommand{\sigmalt}[1]{\sigma_{L,T}^{#1}}

\newcommand{\g}{\gamma}

\newcommand{\csq}{C^2}

\newcommand{\initsig}{\sigma_0/2}

\newcommand{\dof}{\mathrm{dof}}
\newcommand{\chisqdof}{\chi^2/\dof}

\newcommand{\mvgamma}{\mathrm{MV}^{\g}}

\newcommand{\balsd}{\mathrm{Bal+SD}}
\newcommand{\as}{\alpha_\mathrm{s}}

\usepackage[breaklinks,colorlinks,citecolor=citcolor,urlcolor=blue,linkcolor=lcolor]{hyperref}
\definecolor{lcolor}{rgb}{0.5,0,0}
\definecolor{citcolor}{rgb}{0,0.3,0.0}
\definecolor{teal}{rgb}{0.0, 0.5, 0.5}
\begin{document}

\title{Confronting Color Glass Condensate at next-to-leading order with HERA data }

\author{Carlisle Casuga}
\email{carlisle.doc.casuga@jyu.fi}
\affiliation{
Department of Physics, University of Jyväskylä,  P.O. Box 35, 40014 University of Jyväskylä, Finland
}
\affiliation{
Helsinki Institute of Physics, P.O. Box 64, 00014 University of Helsinki, Finland
}
\author{Heikki Mäntysaari}
\email{heikki.mantysaari@jyu.fi}
\affiliation{
Department of Physics, University of Jyväskylä,  P.O. Box 35, 40014 University of Jyväskylä, Finland
}
\affiliation{
Helsinki Institute of Physics, P.O. Box 64, 00014 University of Helsinki, Finland
}

\begin{abstract}
We perform a global analysis of HERA total inclusive cross section and charm quark production data to extract the non-perturbative initial condition for the next-to-leading order Balitsky–Kovchegov (BK) equation. We extend our previous analyses to full next-to-leading order + next-to-leading logarithm (NLO+NLL) accuracy by combining the NLO DIS impact factors with the NLO BK equation that includes the resummation of large double and single logarithms. The developed Bayesian setup extracts the posterior that describes the distribution of the parameters that best fit the data. The posterior distribution allows for estimates of the theoretical uncertainty of the dipole amplitude and, hence, offers a streamlined method for propagating uncertainties in the BK initial condition in NLO CGC calculations.
    
\end{abstract}

\maketitle

\section{Introduction}

The study of the proton’s high-energy structure is both complex and intriguing due to the non-perturbatively large gluon density.
As the collision energy increases, the gluon density grows and eventually reaches the saturation regime, characterized by the scale $Q_s^2$, where non-linear QCD dynamics become dominant. The color glass condensate (CGC)~\cite{Iancu:2003xm} provides a natural framework to describe strong interactions in such dense gluonic systems. Since the saturation scale increases with nuclear mass number, $Q_s^2\sim A^{1/3}$, heavy nuclei offer an ideal system to study saturation effects. 
Consequently, upcoming high-precision nuclear deep inelastic scattering (DIS) measurements at the electron-ion collider (EIC)~\cite{AbdulKhalek:2021gbh}  provide a powerful probe of the saturated regime of the nuclear wave function. 

At high energy, DIS can be described in the dipole picture~\cite{Mueller:1994jq}, where, at leading order,  the incoming virtual photon fluctuates into a quark-antiquark dipole. The dipole interacts eikonally with the gluon-dense target as described within the CGC framework.
The time scale difference between the $\gamma^*\rightarrow q\bar{q}$ splitting and the dipole-target interaction allows for a factorization of the total DIS cross section into a process-dependent hard factor and the universal dipole-target scattering amplitude. The energy (or Bjorken-$x$) dependence of this amplitude is  obtained by solving the Balitsky-Kovchegov (BK)~\cite{Balitsky:1995ub, Kovchegov:1999yj} equation. However, this perturbative evolution equation requires a non-perturbative initial condition that must be inferred from data.

Although clear hints of saturation effects have been seen in various high-energy scattering processes, no conclusive evidence that the saturation regime has been  reached at current collider energies exists~\cite{Morreale:2021pnn}.
Conclusive tests of the saturation picture demand higher precision in CGC calculations. In particular, higher order corrections in $\as$ can be numerically significant, see e.g., Refs.~\cite{Ducloue:2017ftk,Mantysaari:2021ryb,Mantysaari:2022kdm,Kaushik:2025roa,Caucal:2023fsf}. A consistent next-to-leading order (NLO) analysis of DIS should involve both NLO impact factors and next-to-leading log (NLL) corrections to small-$x$ evolution equations. First principle calculations in light-cone perturbation theory for the DIS impact factors at NLO in $\as$ have been performed in Refs.~\cite{Balitsky:2010ze, Beuf:2011xd, Balitsky:2012bs,Beuf:2016wdz, Hanninen:2017ddy, Beuf:2017bpd} for light quarks, and in Refs.~\cite{Beuf:2022ndu,Beuf:2021qqa,Beuf:2021srj} for massive quarks. Similarly, the BK evolution equation at NLL accuracy has been obtained in Ref.~\cite{Balitsky:2007feb} and numerically implemented in Refs.~\cite{Lappi:2015fma, Lappi:2016fmu,Cepila:2024qge}, resumming contributions  enhanced by large logarithms of $x$ (or energy) $\sim \as^2 \ln(1/x)$. 

Quantifying the sources of uncertainty in CGC calculations plays a crucial role in the search for saturation. One major source of model uncertainty is the non-perturbative initial condition of the BK equation, which must be constrained using the precise small-$x$ structure function data~\cite{H1:2015ubc, H1:2018flt}. Parametrizations of the initial condition for the BK equation can be represented by models such as the Mclerran-Venugopalan (MV)~\cite{McLerran:1993ni} or the Golec-Biernat-W{\"u}sthoff (GBW)~\cite{GolecBiernat:1998js} model. Numerous studies, such as Refs.~\cite{Albacete:2009fh, Albacete:2010sy, Lappi:2013zma} at LO and Ref.~\cite{Beuf:2020dxl} for light quarks and Ref.~\cite{Hanninen:2022gje} for heavy quarks at NLO accuracy, extract the non-perturbative initial condition from fits to HERA data. The feasibility of a global fit at NLO is studied in Ref.~\cite{Hanninen:2022gje}, demonstrating the possibility of successfully describing both HERA total reduced cross section~\cite{H1:2015ubc} and heavy quark production~\cite{H1:2018flt} data simultaneously. 

This work extracts the non-perturbative initial condition of the NLO BK equation using the total DIS cross section data and the heavy quark contribution as simultaneous constraints. In the NLO BK equation, we include the resummation of contributions enhanced by large double and single transverse logarithms. As shown in Ref.~\cite{Lappi:2016fmu}, including such resummations in the NLO BK equation is necessary in order to achieve stable evolution. Our previous analysis~\cite{Casuga:2025etc} included only the resummation of large double logarithms by using a kinematically-constrained BK equation (KCBK)~\cite{Beuf:2014uia}, which approximates the NLO BK equation. Using our well-tested Bayesian framework developed in Refs.~\cite{Casuga:2025etc,Casuga:2023dcf}, we obtain posterior distributions estimating the theoretical uncertainty of the initial condition. Samples from the posterior enable straightforward propagation of the initial condition uncertainty to other observables calculated within the CGC framework at NLO accuracy. 

The following section, Sec. \ref{sec:BK}, summarizes the NLO calculation of the inclusive DIS structure functions at NLO, as well as the NLO
BK evolution equation along with the parametrization of its non-perturbative initial condition. In Sec.~\ref{sec:bayesian} we discuss the Bayesian analysis workflow we use to obtain posterior distributions for the model parameters. We follow by discussing our results in Sec.~\ref{sec:results} and conclude in Sec.~\ref{sec:conclusions}.

\section{Deep Inelastic Scattering at Next-to-leading Order}
\label{sec:BK}

We compute the DIS cross section at next-to-leading order accuracy in the dipole picture, comparing results to the precise reduced cross section data from HERA~\cite{H1:2015ubc,H1:2018flt}.
The computational setup is identical to the one used in our previous work~\cite{Casuga:2025etc}, except that in this work we for the first time employ the complete next-to-leading order Balitsky-Kovchegov equation to describe the Bjorken-$x$ dependence of the total DIS cross section. This setup is summarized here for completeness. 

In terms of the proton structure functions $F_2$ and $F_L$, the reduced cross section reads
\begin{equation}
    \sigma_r(y,x,Q^2) = F_2\left(x,Q^2\right) - \frac{y^2}{1+(1-y)^2} F_L\left(x,Q^2\right).
\end{equation}
Here $x$ is the Bjorken-$x$, $y = Q^2/(sx)$ is the inelasticity, and $\sqrt{s}$ is the center-of-mass energy in the electron(positron)-proton scattering. The proton structure functions $F_2$ and $F_L$ are related to the total cross section for the virtual photon-proton scattering $\sigma_{T,L}^{\gamma^* p}$ as
\begin{align}
    F_2 &= \frac{Q^2}{4\pi^2 \aem} (\sigma^{\gamma^* p}_T + \sigma^{\gamma^* p}_L), \\
    F_L &= \frac{Q^2}{4\pi^2 \aem} \sigma^{\gamma^* p}_L.
\label{eq:f2fl}
\end{align}
Here $T$ and $L$ refer to the transverse and longitudinal photon polarization states, respectively.

The total cross section for the photon–proton scattering is computed in the dipole picture using the optical theorem. As such, one computes the forward elastic scattering amplitude for the virtual photon–target scattering, where the photon fluctuates into a partonic state ($|q\bar q\rangle, |q\bar q g\rangle, \dots$) that interacts instantaneously with the target, described as a shockwave. Long after the shockwave, the partonic state recombines into a photon. 
Consequently, the cross section factorizes into a convolution of hard factors describing the partonic subprocess,  and the Wilson line correlators describing how partonic  states interact with the target shockwave.

At leading order, only the $|q\bar q\rangle$ component of the virtual photon contributes to the photon-proton scattering.
The NLO cross section derived in Refs.~\cite{Hanninen:2017ddy,Beuf:2016wdz,Beuf:2017bpd,Beuf:2022ndu,Beuf:2021srj,Beuf:2021qqa} includes virtual loop corrections to the $\gamma\to q\bar q$ splitting, and contributions from processes where the $|q\bar q g\rangle$ component of the photon wave function interacts with the target. The NLO cross section can be written as
\begin{equation}
    \sigmaltnlo = \sigmalt{\textrm{IC}} + \sigmaltdip + \sigmaltqgu .
\end{equation}
We work in the so-called ``unsubtracted scheme''~\cite{Ducloue:2017ftk} where the lowest order term $\sigmalt{\textrm{IC}}$ corresponds to the process where the $|q\bar q\rangle$ component of the virtual photon interacts with the unevolved target (Wilson line correlators evaluated at the initial condition of the BK evolution). Loop corrections to this process are included in the second term
\begin{equation}
\label{eq:sigma_qq}
    \sigmaltdip = K_{q\bar q} \otimes N_{01}.
\end{equation}
The third term corresponds to the case where the $|q\bar qg \rangle $ Fock state of the virtual photon interacts with the target:
\begin{equation}
    \sigmaltqgu = K_{q\bar qg} \otimes N_{012}. 
    \label{eq:sigma_qqg} 
\end{equation}
Explicit expressions for the hard factors for massless quarks can be found from Refs.~\cite{Hanninen:2017ddy,Beuf:2016wdz,Beuf:2017bpd}, and for heavy quarks from Refs.~\cite{Beuf:2022ndu,Beuf:2021srj,Beuf:2021qqa}. In this work, as in Refs.~\cite{Casuga:2025etc,Beuf:2020dxl,Hanninen:2022gje}, we take the three light quark flavors to be massless. The convolution $\otimes$ refers to the integration over the quark transverse coordinates and longitudinal momentum fractions. A publicly available numerical implementation of the NLO DIS cross section (supporting only quarks with non-zero mass) is available at~\cite{Hanninen:2026rbd}. Exploring the sensitivity of our results on the light quark mass is left for future work. We, however, note that previous leading order fits have found only a modest dependence on this mass~\cite{Mantysaari:2018nng,Albacete:2010sy}.

The Wilson line correlators read as
\begin{align}
     S_{01} &= \frac{1}{\nc} \left\langle  \Tr{V(\xt_0) V^\dagger(\xt_1)} \right\rangle , \label{eq:s01} \\
        S_{012} &= \frac{\nc}{2\cf} \left( S_{02}S_{21} - \frac{1}{\nc^2} S_{01}\right). \label{eq:s012}
\end{align}
One can define from here the amplitudes $N_{01}=1-S_{01}$ and $N_{012}=1-S_{012}$. 
The $V(\xt)$ is a Wilson line in the fundamental representation, describing the eikonal interaction of a quark at transverse coordinate $\xt$ when it propagates in the target color field.
Furthermore, $\xij{0}$ and $\xij{1}$ are the quark and antiquark transverse coordinates, and $\xt_2$ is the gluon coordinate. The average $\langle \mathcal{O}\rangle$ refers to an average over target configurations. 
This work forgoes the impact parameter $\bt$ dependence of the dipole amplitude and replaces it with a constant proton transverse area denoted by $\sigma_0/2$.
This is considered as a free parameter of the model to be determined from DIS data\footnote{While such factorized impact parameter dependence is a typical approximation in dipole picture fits~\cite{Lappi:2013zma,Albacete:2010sy,Casuga:2023dcf,Casuga:2025etc}, we also note that finite-size effects may become numerically important in the case of proton targets as  argued e.g. in Refs.~\cite{Berger:2011ew,Mantysaari:2018zdd,Mantysaari:2024zxq}.}.

The energy or Bjorken-$x$ dependence of the Wilson line correlators is given by the Balitsky-Kovchegov equation~\cite{Kovchegov:1999yj,Balitsky:1995ub}. 
The BK equation is an integro-differential equation that we initialize with a McLerran-Venugopalan (MV) model~\cite{McLerran:1993ni} inspired parametrization (used e.g. in Refs.~\cite{Albacete:2010sy, Lappi:2013zma,Beuf:2020dxl,Casuga:2023dcf,Casuga:2025etc}):
\begin{multline}
    \label{eq:bk-ic}
    S_{01}(Y=0)  =
    \exp 
        \left[ 
            -\frac{\left(\xij{01}^2Q_{s,0}^2\right)^\gamma}{4} \right.
             \\
            \left. \times \ln \left( \frac{1}{|\xij{01}| \Lambda_\text{QCD}}
             +   e   \right)   \right]. 
\end{multline}
where $Y$ is the evolution rapidity and the parameter $\qso^2$ parametrizes the proton saturation scale at the initial condition of the evolution. 
Here we use the notation $\xij{ij}=\xij{i}-\xij{j}$. The other free parameter is the anomalous dimension, $\g$. Moving forward following the nomenclature from Ref. ~\cite{Lappi:2013zma}, we will call this setup where $\g$ is made a free parameter the $\mvgamma$ initial condition. In the $\sigmaltdip$ term, the Wilson lines are evaluated at the evolution rapidity $Y=\ln 1/x$. When evaluating the $\sigmaltqgu$ contribution, $Y=\ln W^2 z_2/Q_0^2$ where $z_2$ is the fraction of the photon's large light-cone plus momentum carried by the gluon (that one intergrates over), $W$ is the center-of-mass energy of the photon-proton system, and $Q_0^2$ is a non-perturbative momentum scale that  we set to $Q_0^2=1\,\mathrm{GeV}^2$. For a more detailed discussion, see Refs.~\cite{Casuga:2025etc,Beuf:2020dxl}.

The free parameters in the initial dipole-proton scattering amplitude in Eq. \eqref{eq:bk-ic} are determined by performing a Bayesian analysis as discussed in Sec.~\ref{sec:bayesian}. In this work, similarly as in our previous work~\cite{Casuga:2023dcf}, we consider two initial condition parametrizations. In the baseline setup, we use the MV model dipole where one sets $\g=1$. Then, we consider a more flexible initial condition where $\g$ is let to be a free parameter. We note that in previous leading~\cite{Albacete:2010sy,Lappi:2013zma} and next-to-leading~\cite{Beuf:2020dxl,Casuga:2023dcf,Casuga:2025etc} order fits a large $\g>1$ has been required in order to obtain a good description of the HERA data.

The NLO BK equation derived in Ref.~\cite{Balitsky:2007feb} reads as

\begin{align}
\label{eq:nlobk}
\begin{split}
    \partial_Y S_{01} = \frac{\as N_c}{2\pi^2} K_1 &\otimes [S_{02}S_{12}-S_{01}] \\ 
    + \frac{\as^2 N_c^2}{8\pi^4}  K_2 &\otimes [S_{02}S_{22'}
    S_{12'}
    - S_{02}S_{12}] \\ 
    + \frac{\as^2 \nf N_c}{8\pi^4} K_f  &\otimes S_{12}[S_{02'} 
    -S_{01}] .
\end{split}
\end{align}
Explicit expressions for the hard factors $K_1, K_2$ and $K_f$ can be found from Refs.~\cite{Balitsky:2007feb,Lappi:2015fma}. In this work we include, as in Ref.~\cite{Casuga:2025etc}, three massless light quark flavors and a charm quark when calculating the DIS cross section, and consequently set $\nf=4$ on the third line in Eq.~\eqref{eq:nlobk}. The convolution $\otimes$ corresponds to an integration over the transverse coordinates.  

As explicitly demonstrated in Refs.~\cite{Lappi:2015fma,Lappi:2016fmu}, stabilizing the NLO BK equation requires one to resum contributions enhanced by large transverse logarithms $\sim \ln (\xij{02}^2/\xij{01}^2) \ln (\xij{12}^2/\xij{01}^2)$. As shown in Ref.~\cite{Iancu:2015vea}, such contributions can be resummed by multiplying the kernel $K_1$ by the factor
\begin{equation}
     K_{\mathrm{DLA}} = \frac{J_1\left( 2 \sqrt{\bar{\as}x^2} \right)}{\sqrt{\bar{\as}x^2}} 
\end{equation}
and removing the explicit double logarithmic term from $K_1$ obtained in Ref.~\cite{Balitsky:2007feb}. 
Here $x=\ln (\xij{02}^2/\xij{01}^2) \ln (\xij{12}^2/\xij{01}^2)$ and $ \bar{\as}= \as \nc/\pi$. 
If $\ln X^2/r^2 \ln Y^2/r^2<0$,  an absolute value is used and the Bessel function is changed from $J_1$ to $I_1$.
Including the resummation of these double logarithmic contributions in the leading order BK equation is parametrically equivalent to the kinematically constrained BK equation (KCBK)~\cite{Beuf:2014uia} used in our previous analysis~\cite{Casuga:2025etc} to approximate the NLO BK evolution, where one imposes a time ordering between consecutive gluon emissions.

Single transverse logarithms enhanced for small dipole sizes can also be resummed to all orders in $\as$. Following Ref.~\cite{Iancu:2015joa}, this resummation is done by including the factor 
\begin{equation}
    K_{\mathrm{STL}} = \exp \left\{ - \frac{\as \nc A_1}{\pi} \left| \ln \frac{C_{\mathrm{sub}}\xij{01}^2}{\mathrm{min}[\xij{02}^2,\xij{12}^2]} \right| \right\}
\end{equation}
in $K_1$.
Here $A_1 = 11/12$, and $C_\mathrm{sub} = 0.65$ is determined in Ref.~\cite{Lappi:2016fmu} to minimize the remaining $\as^2$ corrections in the NLO BK evolution that are not included in the resummations. With the single and double logarithmic corrections resummed, the hard factor $K_1$ can be written as
\begin{multline}
    \frac{\as \nc}{2\pi^2} K_1 \to  \frac{\as(\xij{01}^2)\nc}{2\pi^2} K_\mathrm{DLA} K_\mathrm{STL} \biggl[ \frac{\xij{01}^2}{\xij{02}^2\xij{12}^2} \\ + \frac{1}{\xij{02}^2} \biggl( \frac{\as(\xij{02}^2)}{\as(\xij{12}^2)}  - 1\biggr) + \frac{1}{\xij{12}^2} \biggl( \frac{\as(\xij{12}^2)}{\as(\xij{02}^2)}  - 1\biggr) \biggr] \\ - K_\mathrm{sub} + K_1^\mathrm{fin},
\label{eq:finalK1}
\end{multline}
where $K_\mathrm{sub}$ removes the $\mathcal{O}(\as^2)$ contribution of $K_\mathrm{STL}$ that appears in $K_2$ in exact kinematics. Meanwhile, $K_1^\mathrm{fin}$ contains the remaining finite $\as^2$ terms that are not enhanced by large transverse logarithms. See Ref.~\cite{Lappi:2016fmu} for explicit expressions.

In Eq.~\eqref{eq:finalK1}, we have used the Balitsky prescription for the running coupling in the BK equation~\cite{Balitsky:2006wa}. We will refer to this running coupling prescription as $\balsd$ in this work. 
As in this prescription the smallest dipole size parametrically controls the scale of the running coupling, we also choose to evaluate all other factors $\as$ in the BK equation and in the DIS impact factors at the scale set by the smallest dipole size: $\as = \as(\mathrm{min}\{\xij{01}^2, \xij{02}^2, \xij{21}^2 \})$. Similarly as in Ref. \cite{Casuga:2025etc}, we study sensitivity to the running coupling prescription by performing one analysis using a   prescription where the parent dipole size is always used to set the scale of the running coupling: $\as = \as(\xij{01}^2)$.

The coordinate space running coupling in the BK evolution is evaluated as
\begin{equation}
\label{eq:running_coupling_vfs}
    \as(\xij{ij}^2) = \frac{4\pi}{\beta_{0, \nf}\log \left( \frac{4C^2}{\xij{ij}^2\Lambda_{\nf}^2} \right)}
\end{equation}
where $\beta_{0, \nf} = (11\nc-2\nf)/3$. We parametrize the scale of the strong coupling in the coordinate space with a free parameter, $C^2$, that dictates the evolution speed and also accounts for the corrections from higher order contributions. 

The number of active quark flavors $\nf$ in Eq.~\eqref{eq:running_coupling_vfs} is chosen according to the variable flavor scheme following Ref.~\cite{Albacete:2010sy} used to account for the heavy quark contribution to the running coupling. It is set according to the number of quark flavors that are lighter than the momentum scale associated with the distance scale, $\xij{ij}^2$, at which $\as$ is evaluated. The $\nf$ dependence of the reference scale $\Lambda_{\nf}$ 
is obtained according to a matching condition, requiring that in the branches of adjacent $\nf$ the strong coupling is continuous: $\alpha_{s,\nf-1}(4C^2/m_\mathrm{f}^2) = \alpha_{s,\nf}(4C^2/m_\mathrm{f}^2)$. This results to 
\begin{equation}
    \label{eq:lqcd}
    \Lambda_{\nf-1} = m_\mathrm{f}^{1-\frac{\beta_{0,\nf}}{\beta_{0,\nf-1}}}\ \Lambda_{\nf}^{\frac{\beta_{0,\nf}}{\beta_{0,\nf-1}}}.
\end{equation}
The reference scale  $\Lambda_5$ is obtained in Ref.~\cite{Albacete:2010sy} by using the $Z^0$ mass, $91.1876 \ \gev$, as a reference requiring $\as(m^2_{Z^0})=0.1184$ in the case where five quark flavors are active. This gives $\Lambda_{\nf=5}=0.0904389\,\mathrm{GeV}$. Although we keep the charm quark mass as a free parameter when calculating DIS cross sections, when determining $\nf$ we use constant quark masses $m_c=1.3\,\mathrm{GeV}$ and $m_b=4.5~\mathrm{GeV}$. In the DIS impact factors, on the other hand, we use exactly the same $\nf$-independent running coupling prescription as in our previous fit~\cite{Casuga:2025etc}.

\section{Bayesian Inference Framework}
\label{sec:bayesian}

The framework used to constrain the non-perturbative parameters is the same as in our previous analyses. We provide a brief summary below,  further details can be found from Refs.~\cite{Casuga:2023dcf, Casuga:2025etc}.

The basic structure of the Bayesian framework is the combination of a trained Gaussian process (GP) emulator~\cite{scikit-learn} and a Markov Chain Monte Carlo (MCMC) sampling algorithm. The MCMC algorithm scans the bounded parameter space to sample the probability distribution of best-fit parameter vectors. The emulator learns the parameter dependence of the model as it is trained with model calculations (total reduced cross section and the charm quark contribution). The calibrated emulator becomes a faster model substitute for the MCMC algorithm to use. The \texttt{emcee} ~\cite{emcee} implementation of the MCMC involves an ensemble of random walkers whose advancements depend on information from other walkers. A walker proposes their next step and is accepted with a probability similar to that of a Metropolis-Hastings algorithm: $\alpha \sim P(\btheta_1) / P(\btheta_0) $, where $\btheta_0$ is the walker's current position in the parameter space,  $\btheta_1$ is the proposed step and $P$ is the likelihood. The walkers converge to regions of high posterior probability, and their chain comprise the posterior samples. In order to achieve higher precision in the physically most relevant part of the parameter space, we perform iterative training. We start with a broad prior range for the model parameters, perform the Bayesian analysis, and then train the emulators again in the region of phase space where the likelihood is found to be large. 

\begin{figure}[t]
    \centering
    \subfloat[MV, Balitsky+smallest dipole]{
    \includegraphics[width=0.48\linewidth]{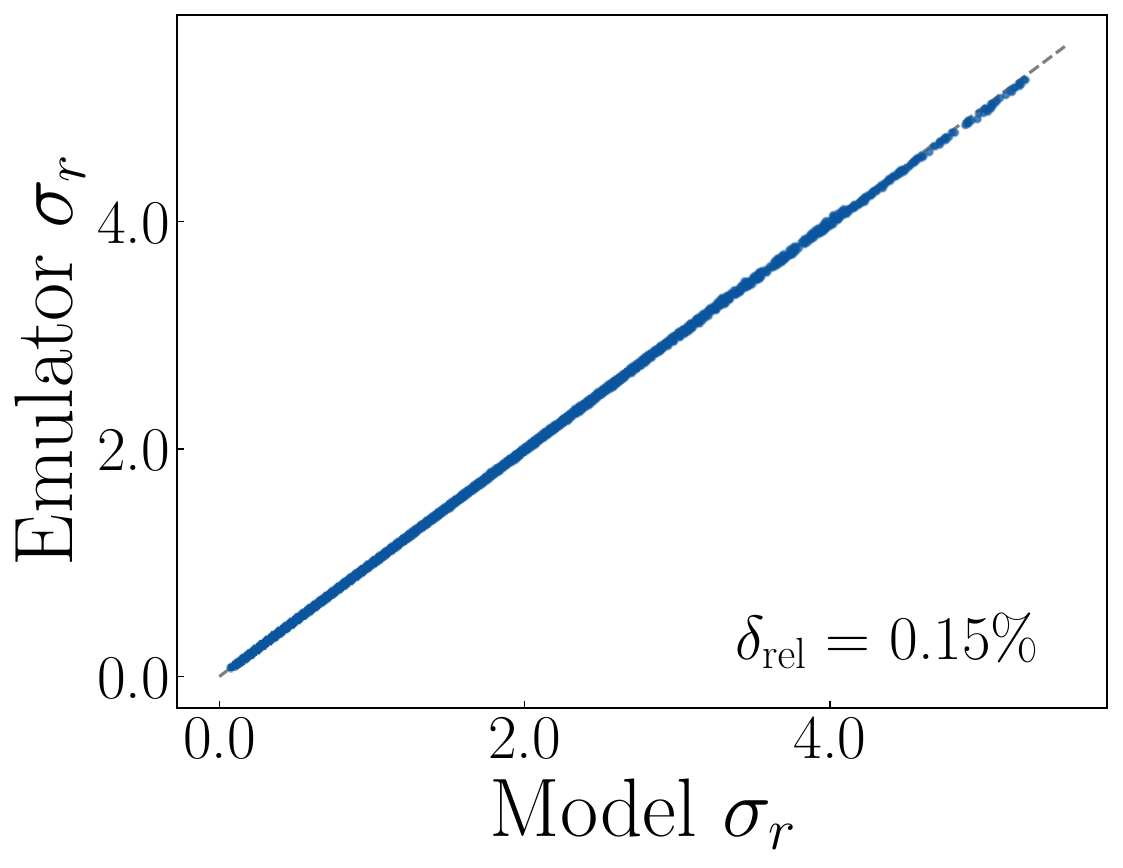}
    \includegraphics[width=0.48\linewidth]{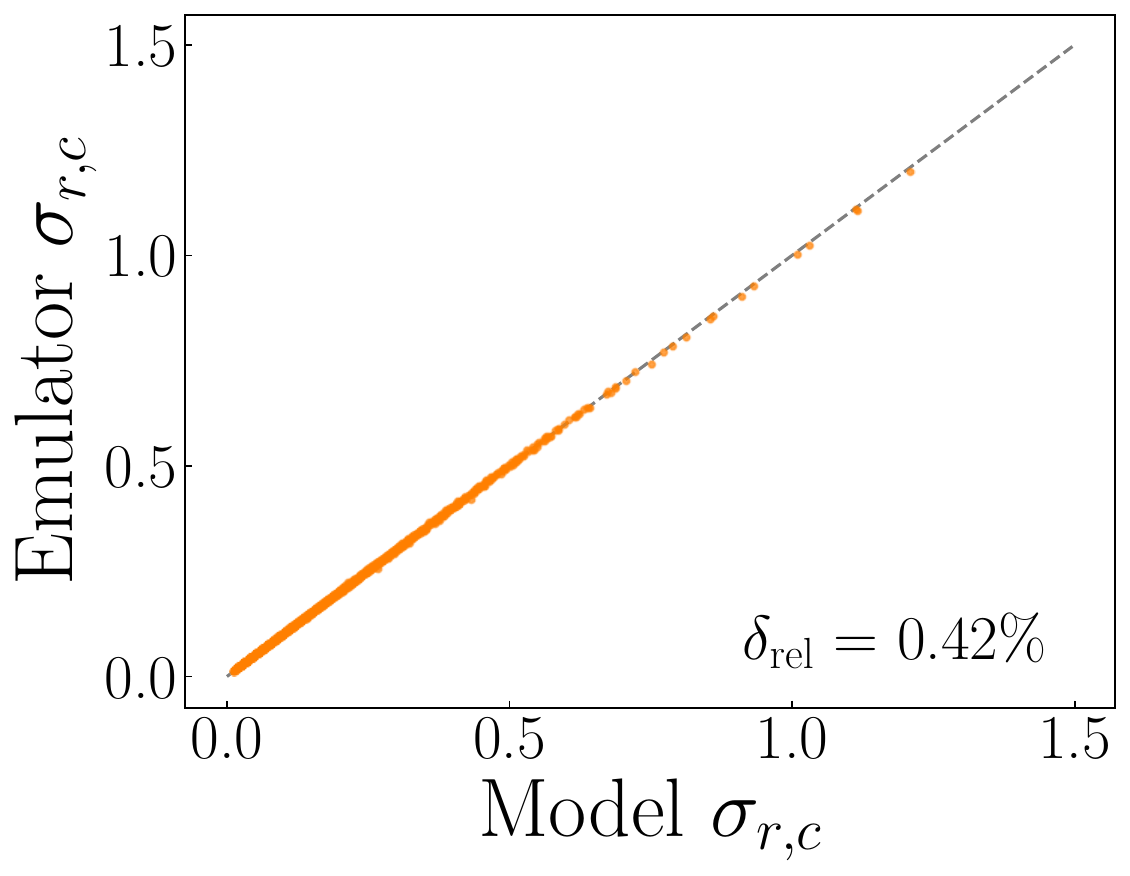}
    \label{fig:training_mv}
    } \\
    \subfloat[$\mvgamma$, Balitsky+smallest dipole]{
    \includegraphics[width=0.48\linewidth]{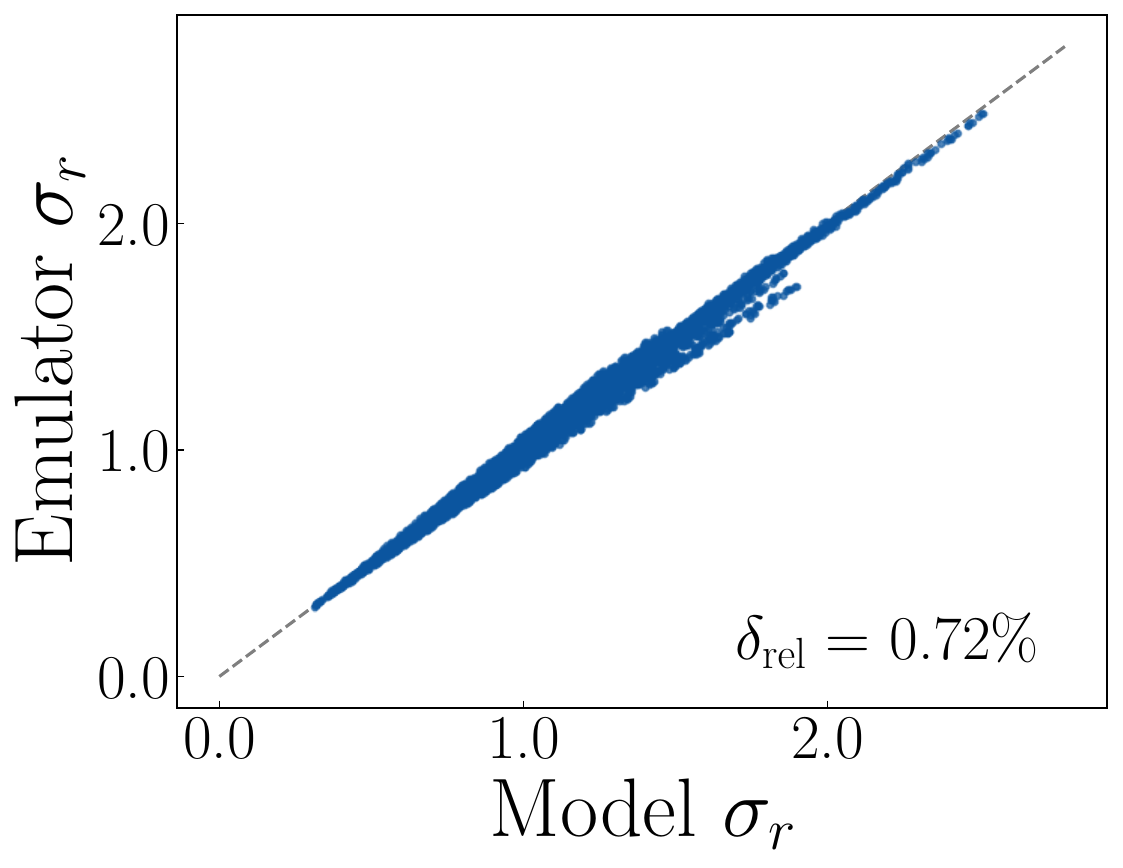}
    \includegraphics[width=0.48\linewidth]{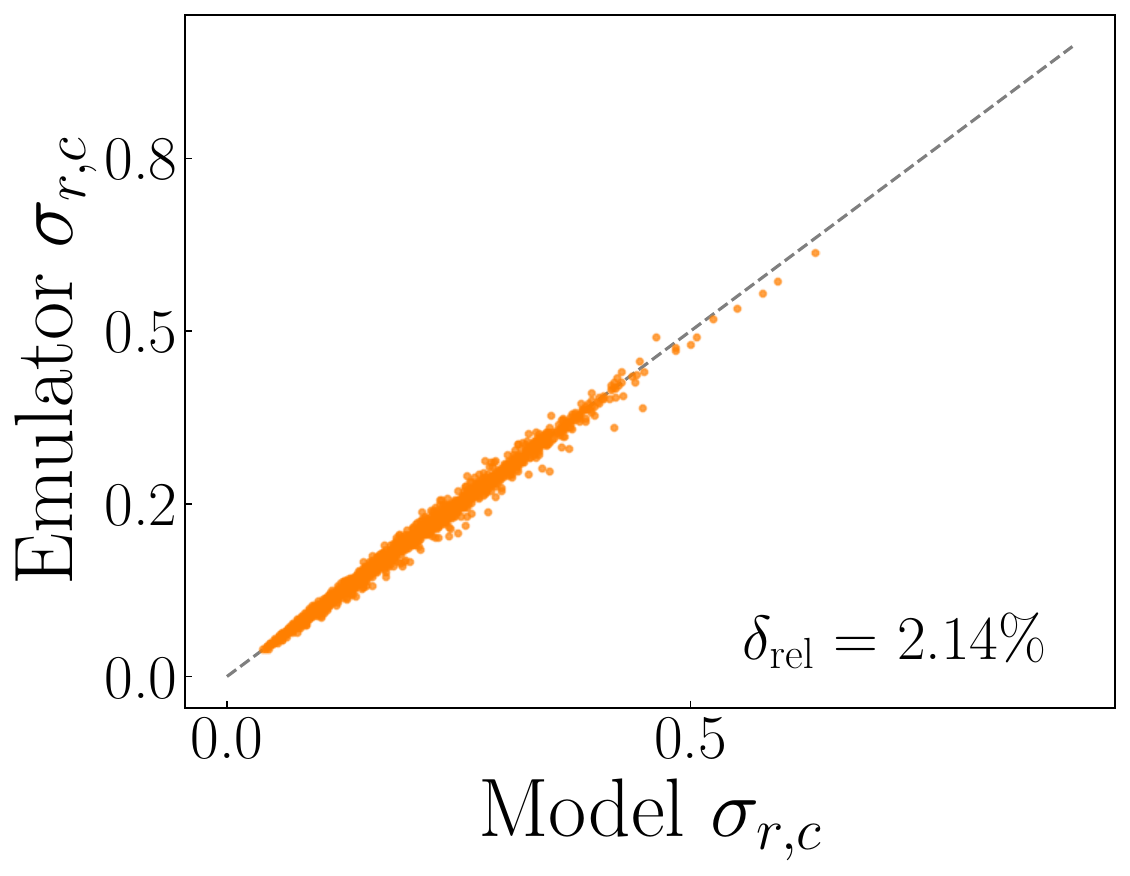}
    \label{fig:training_balsd}
    } \\
    \subfloat[$\mvgamma$, Parent dipole]{
    \includegraphics[width=0.48\linewidth]{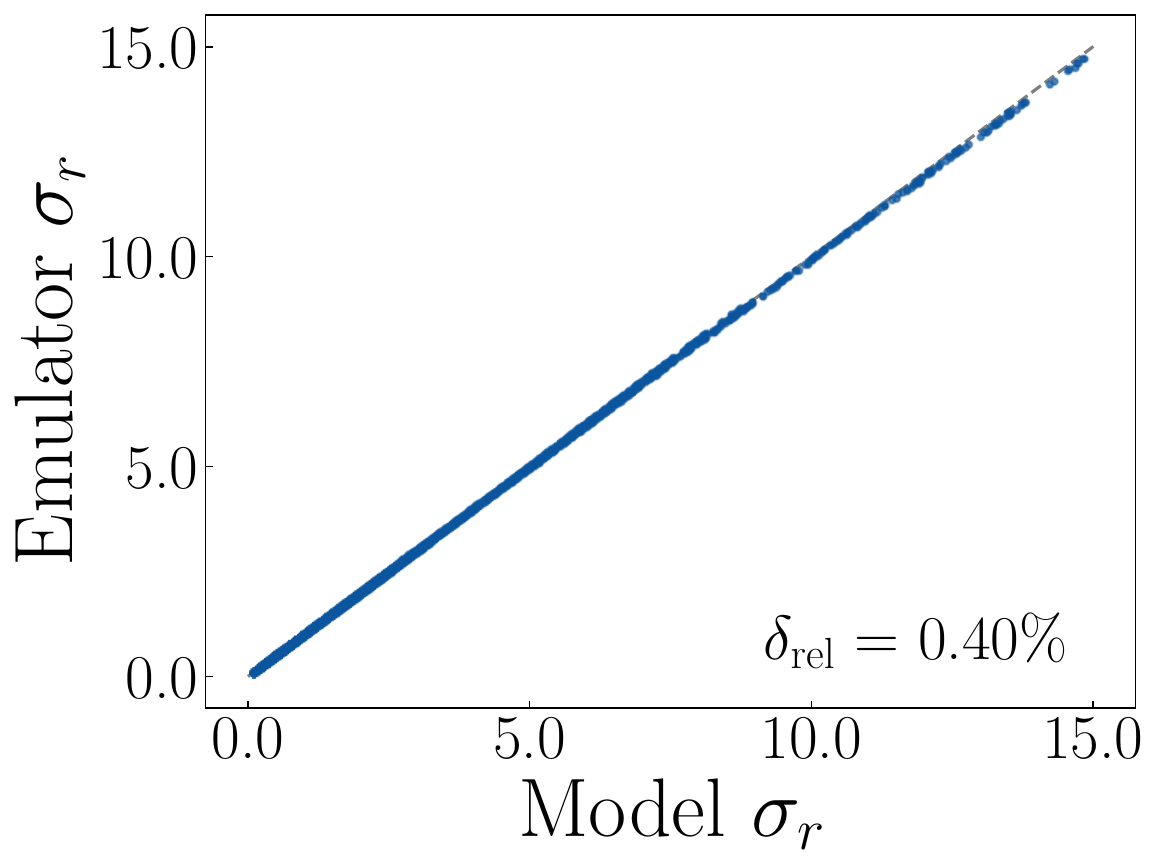}
    \includegraphics[width=0.48\linewidth]{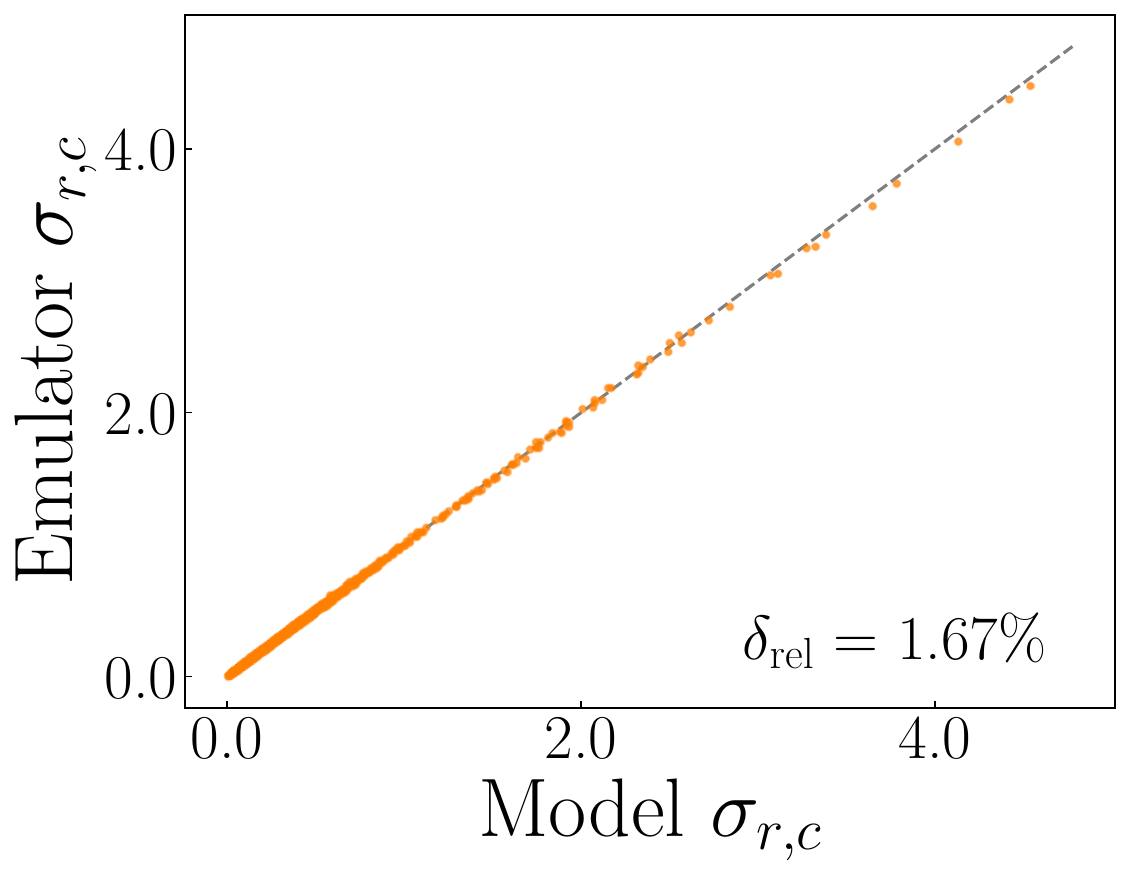}
    \label{fig:training_pd}
    }
    \caption{Comparison between emulator prediction and model calculation for the total reduced cross section (left) and charm contribution (right). The relative difference averaged over all training samples is presented.}
    \label{fig:training}
\end{figure}

\begin{figure*}[t]
    \centering
    \subfloat[MV, Balitsky+smallest dipole]{
    \includegraphics[width=0.3\linewidth]{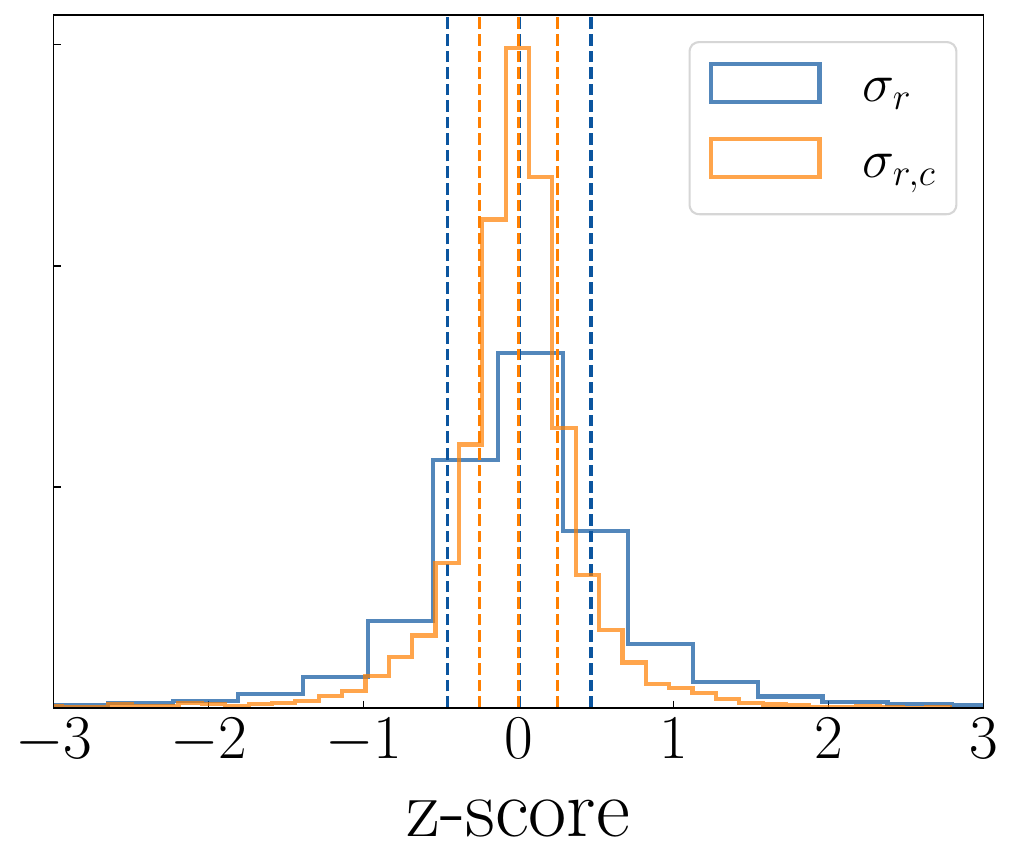}
    }
    \subfloat[$\mvgamma$, Balitsky+smallest dipole]{
    \includegraphics[width=0.3\linewidth]{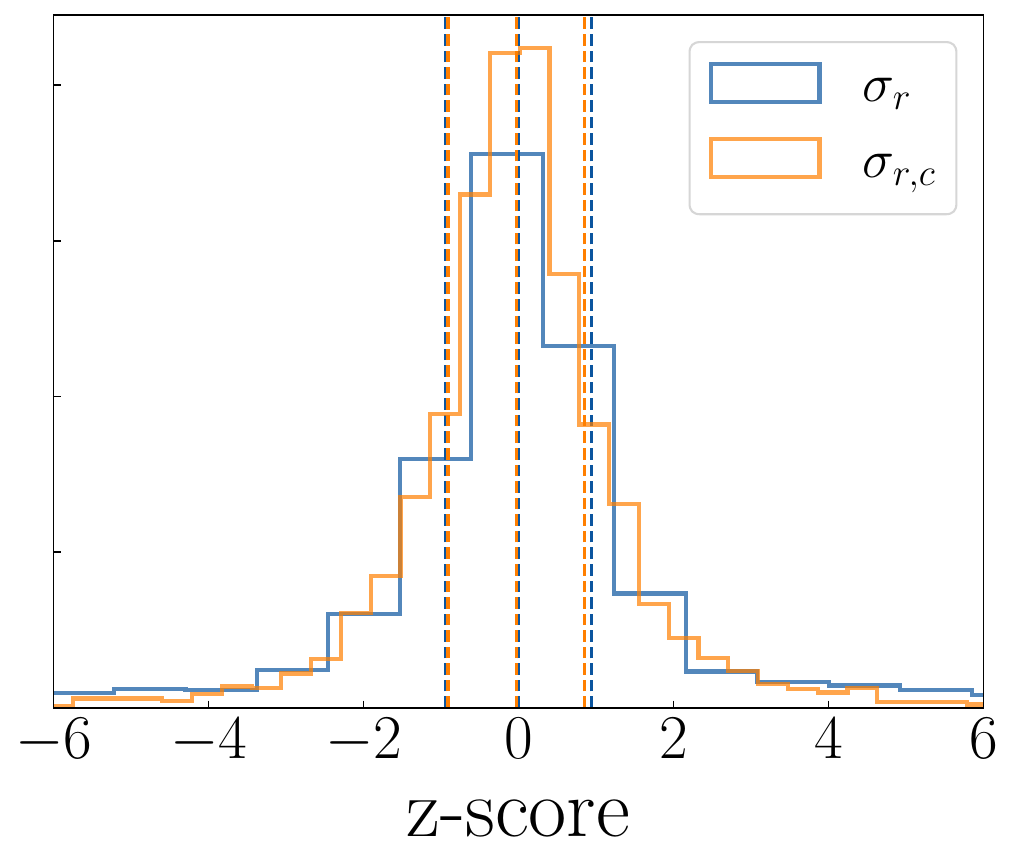}
    }
    \subfloat[$\mvgamma$, Parent dipole]{
    \includegraphics[width=0.3\linewidth]{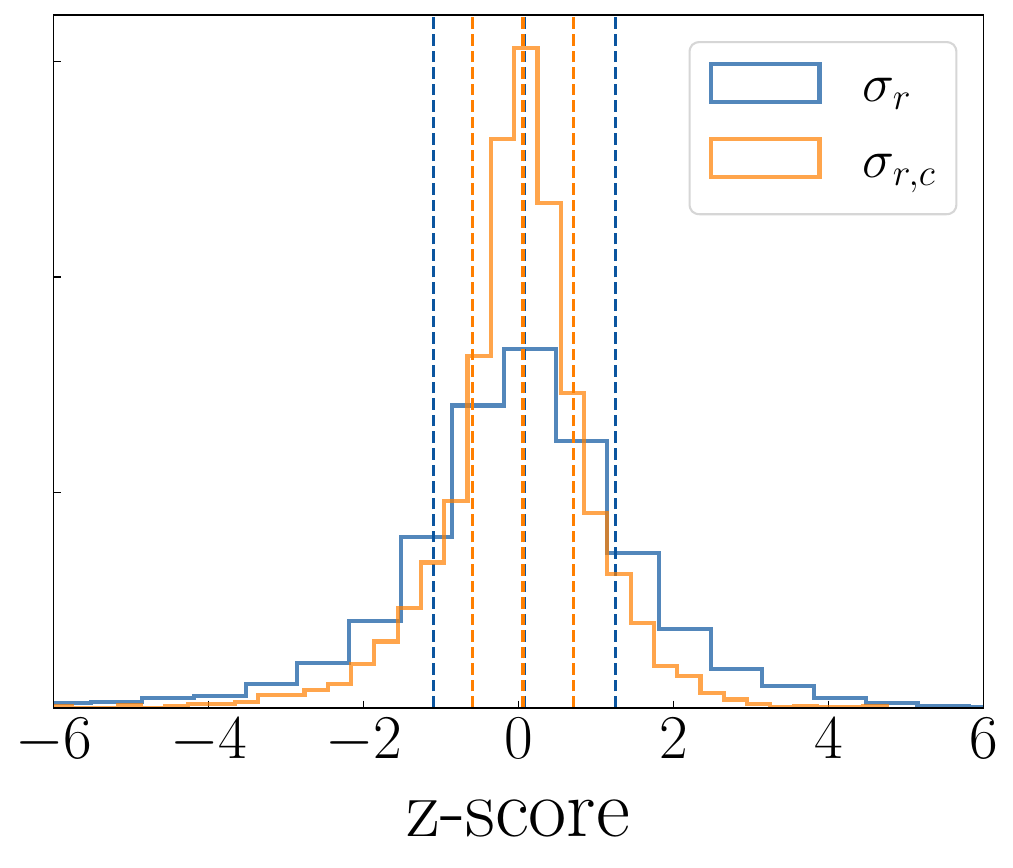}
    }
    \caption{Z-score distribution, vertical lines correspond to the median and one standard deviation of the distribution}
    \label{fig:zscore}
\end{figure*}

The GP emulator uses a maximum likelihood estimator to optimize the hyperparameters of a Gaussian kernel to model the parameter dependence of a function. It pairs seamlessly with the MCMC algorithm because the GP is able to estimate its predictive correlated uncertainty. The GP is trained  over $\sim500-750$ design points obtained to form a Latin-hypercube grid. The training data are further reduced to $4-6$ principal components covering $> 99\%$ of the variance. Correlated systematic uncertainties in the constraining HERA structure function data are taken into account when computing the likelihood as in Refs.~\cite{Casuga:2023dcf, Casuga:2023dcf}.

We first validate our GPEs by comparing the emulator prediction to the corresponding model calculation. This comparison is shown separately for the three different setups applied in this work. First, in Fig.~\ref{fig:training_mv} we use the MV model initial condition (setting $\g=1$ in Eq.~\eqref{eq:bk-ic}), and  the Bal+SD running coupling prescription. Corresponding results for the MV$^\gamma$ initial condition, obtained using the Bal+SD or the parent dipole running coupling prescription, are shown in Figs.~\ref{fig:training_balsd} and~\ref{fig:training_pd}, respectively.
In these figures we also show the average relative error. In all cases an excellent agreement is obtained especially in the case of the total reduced cross section. In the charm quark production the emulator accuracy is slightly worse. This is because the total cross section depends only weakly on the charm quark mass, and consequently the parameter space is effectively larger in the case of charm production. The parameter $\initsig$, on which the $\gamma$-proton cross sections depends linearly, can be excluded from the GPE training to reduce emulator error.

The $z$-score distribution in Fig.~\ref{fig:zscore} evaluates the robustness of the emulator uncertainty estimate.  We show the distribution of $z = (y_\mathrm{GPE} - y_\mathrm{model})/\sigma_\mathrm{GPE}$, where $y_\mathrm{GPE}$ and $y_\mathrm{model}$ are the GPE prediction and the model result, respectively, and $\sigma_\mathrm{GPE}$ is the uncertainty estimated by the GPE. This distribution is again shown separately for the three different setups studied in this work. A $z$-score distribution with unit width and zero mean indicates that the emulator is able to quantify the uncertainty appropriately. All distributions are accurately peaked at zero, and have a width $\sim 1$, implying that the uncertainty estimates are reliable.

\begin{figure*}[t]
    \centering
    \subfloat[MV, Balitsky+smallest dipole]{
    \includegraphics[width=0.8\linewidth]{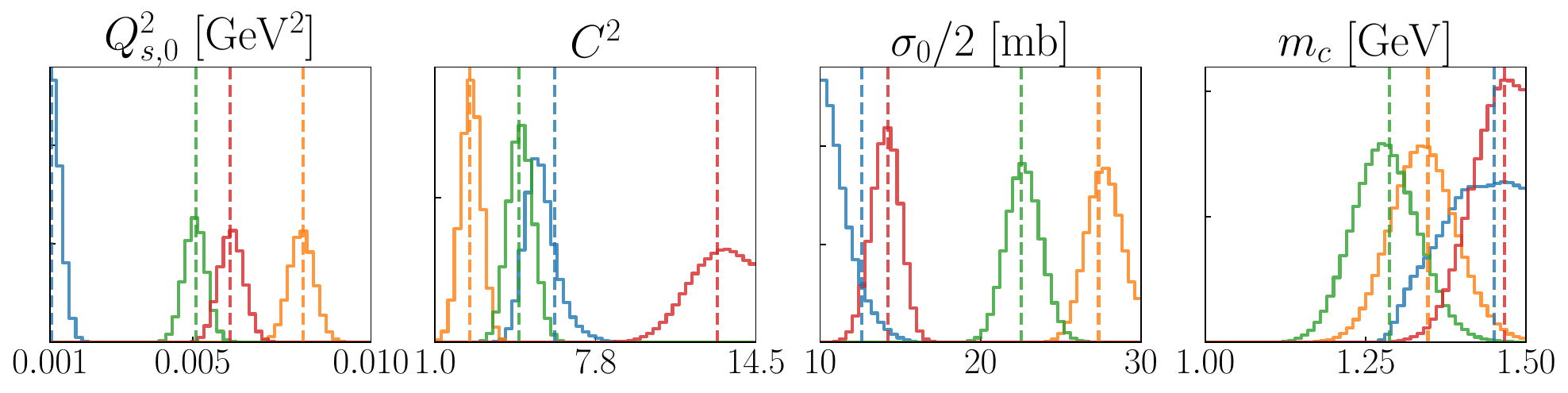}}
    \\
    \subfloat[$\mvgamma$, Balitsky+smallest dipole]{
    \includegraphics[width=\linewidth]{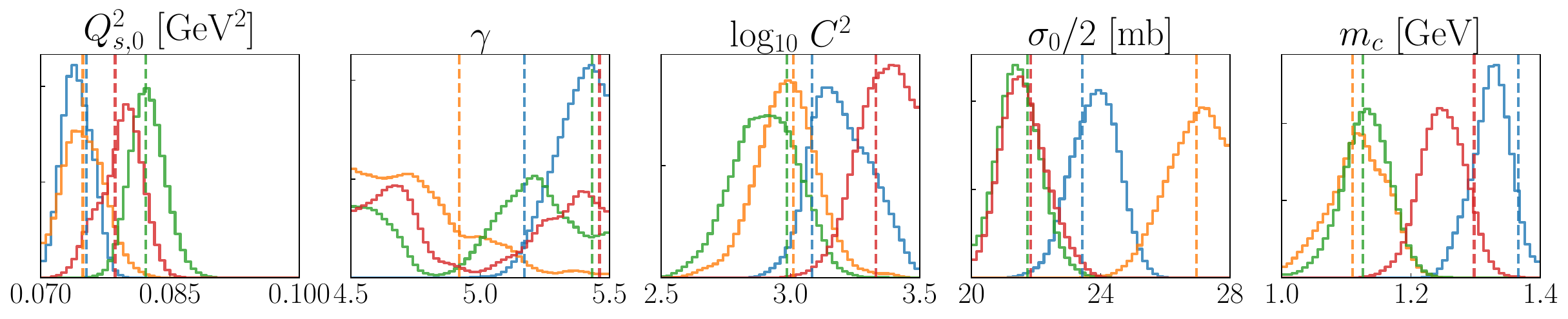}}
    \\
    \subfloat[$\mvgamma$, parent dipole]{
    \includegraphics[width=\linewidth]{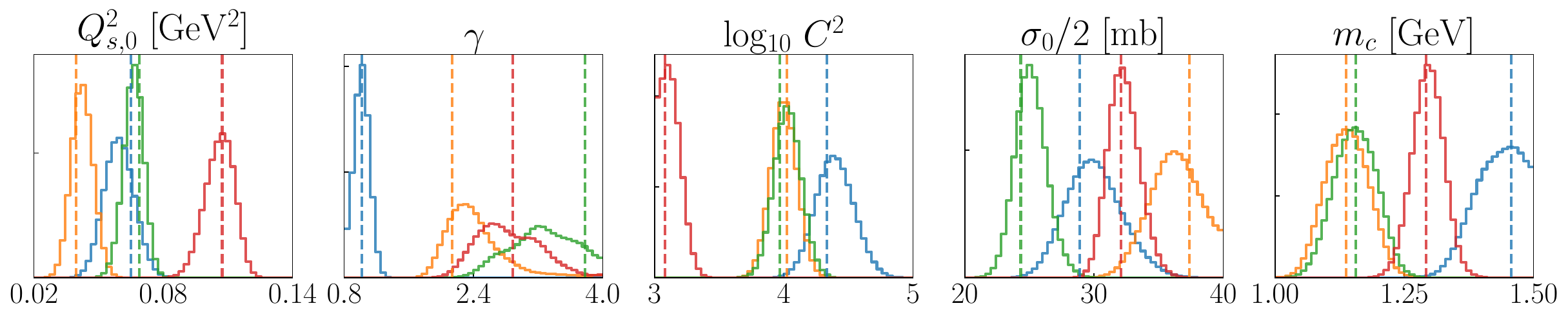}}
    \caption{One-dimensional projections of posterior distributions from closure tests for the Bayesian setup, constraining on pseudodata generated from known parameters (dashed vertical lines) 
    }
    \label{fig:closuretests}
\end{figure*}

We also perform a closure test over the final prior space to maintain our confidence in our Bayesian setup. In a closure test, pseudodata are generated from a random but known parameter vector (truth) excluded from the emulator training set. When constraining our model with the pseudodata, one should recover the said true parameters. Fig.~\ref{fig:closuretests} shows the 1-dimensional projections of different posterior distributions using the three different setups considered in this work. The vertical lines indicate the corresponding true parameter values.

Ideally, the resulting posterior distributions should converge around the truth values. This is the case when using the MV initial condition where the posterior distributions are well-peaked, except for the $\csq$ parameter to which the initial condition depends logarithmically and consequently relatively broad distributions are obtained. For the $\mvgamma$ case, most parameters converge well to the true values except for the $\g$ parameter especially when the $\balsd$ coupling is used. This is because in the closure test we focus on the  region of  high posterior distribution that we will obtain in Sec.~\ref{sec:results}, and when the $\balsd$ prescription is used, the preferred values for the anomalous dimension are large.  
With a large $\g$, the dipole amplitude at the initial condition resembles a step function, and the BK evolution will rapidly develop a small-$r$ tail for the dipole that does not depend strongly on the initial $\g$. Consequently, it is computationally challenging to distinguish initial conditions with different anomalous dimensions.

In this work we extract the posterior distribution of the 5-dimensional parameter vector, ~$\btheta= \{\qso^2, \ \g, \ \log_{10}C^2, \ \initsig,\ m_c\}$, using the total reduced cross section ~\cite{H1:2015ubc} and charm quark contribution ~\cite{H1:2018flt} data measured in deep elastic $ep$ scattering from HERA as constraints. Similar to our previous analyses, we consider data in the small-$x$ region, $x \leq0.01$, with a kinematical cut at $2 \ \gev^2 \leq Q^2 \leq 45 \ \gev^2 $. 

\section{Results}
\label{sec:results}

We first show fit results for different initial condition parametrizations and running coupling prescriptions. 
Results for the $F_2$ structure function outside in the kinematical region that goes beyond the HERA coverage, as well as predictions for $F_L$ that is not included in the fit, are shown in Sec.~\ref{sec:predictions}.
Parameter vectors sampled from the posterior distribution, as well as the corresponding NLO BK-evolved dipole amplitudes, are available at~\cite{samples_zenodo_nlobk}.

\subsection{MV dipole initial condition}

\begin{figure*}
    \centering
    \includegraphics[width=0.77\linewidth]{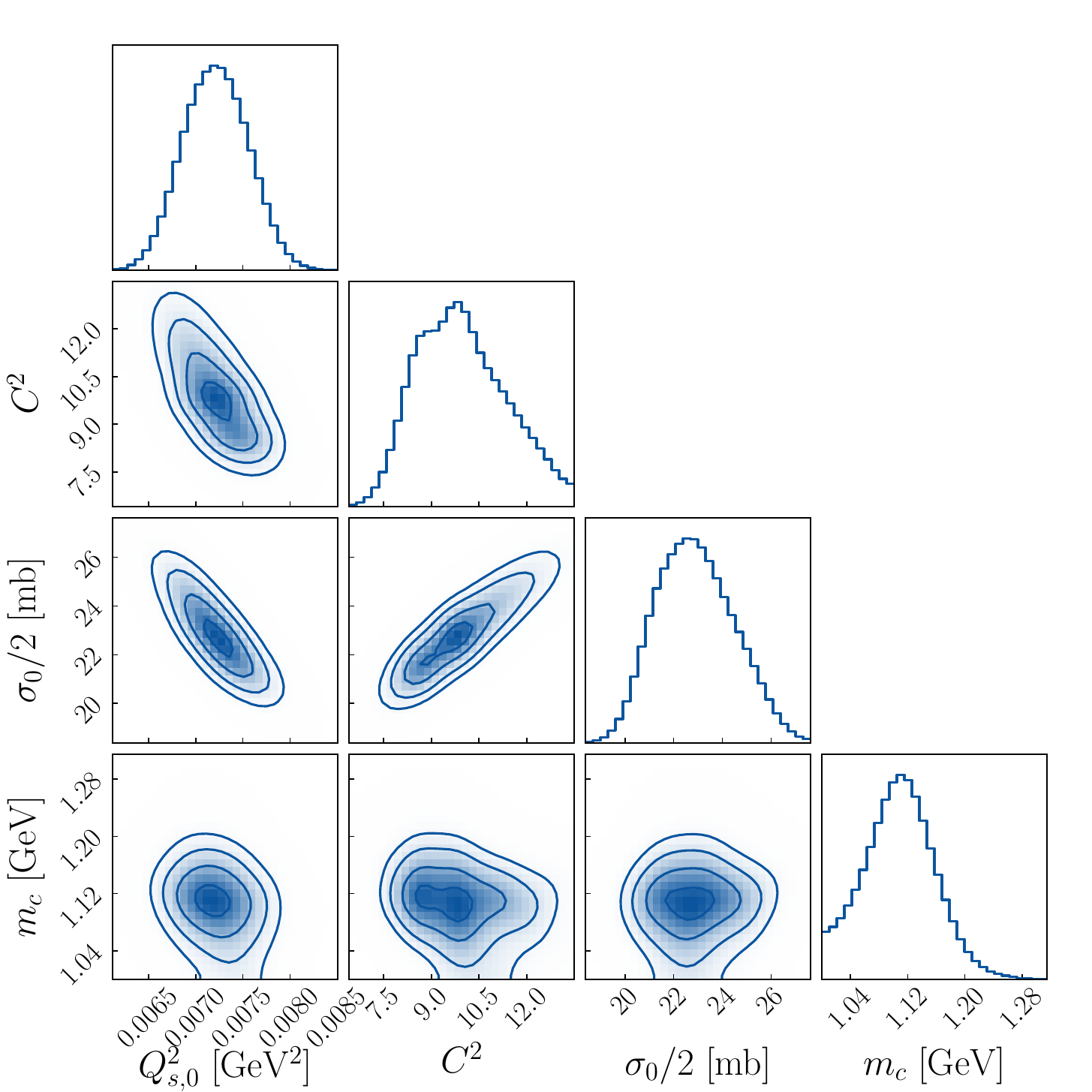}
    \caption{1- (diagonal) and 2- (off-diagonal) projections of the 4-dimensional posterior distribution obtained for the MV model initial condition (fixing $\g=1$) using the $\balsd$ running coupling prescription. }
    \label{fig:mv_balsd}
\end{figure*}

\begin{figure*}[tb]
    \centering
    \includegraphics[width=0.49\linewidth]{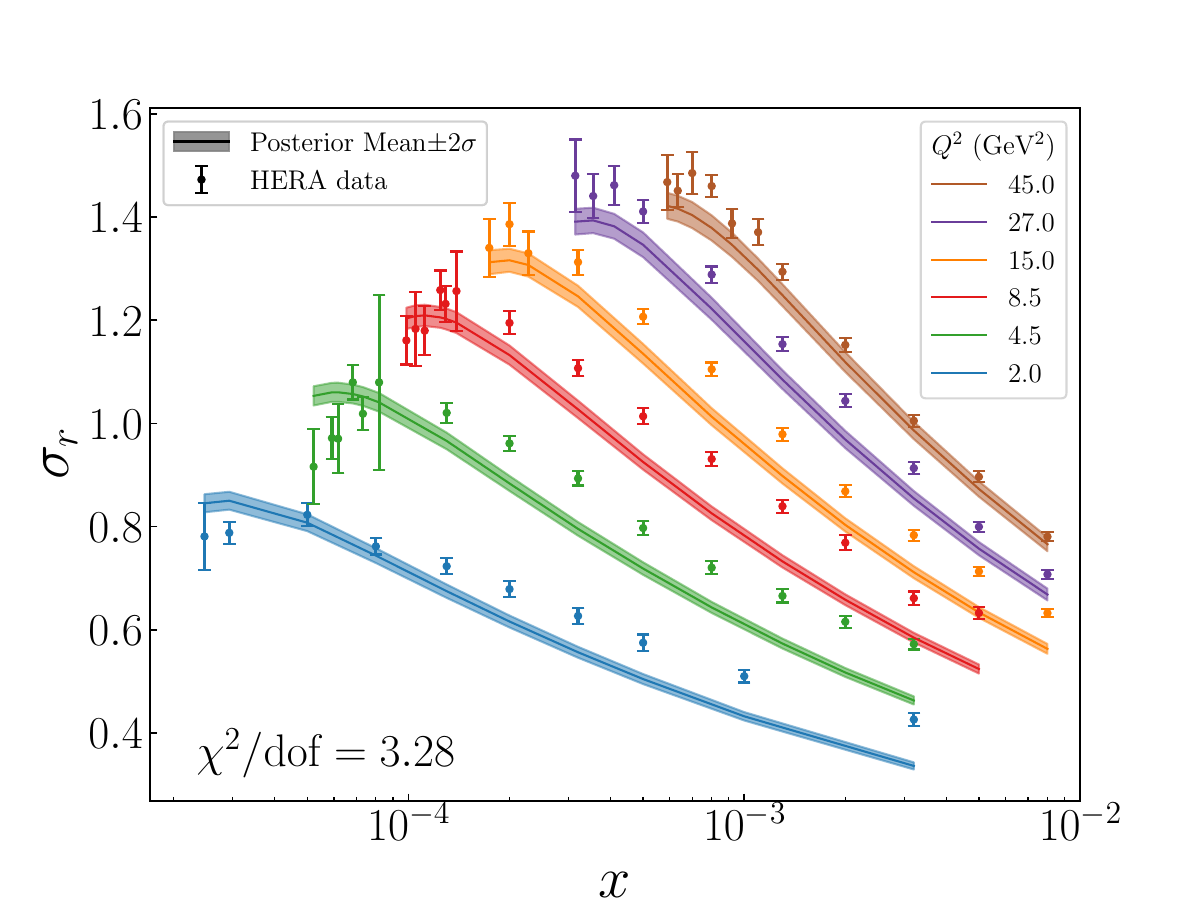}
    \includegraphics[width=0.49\linewidth]{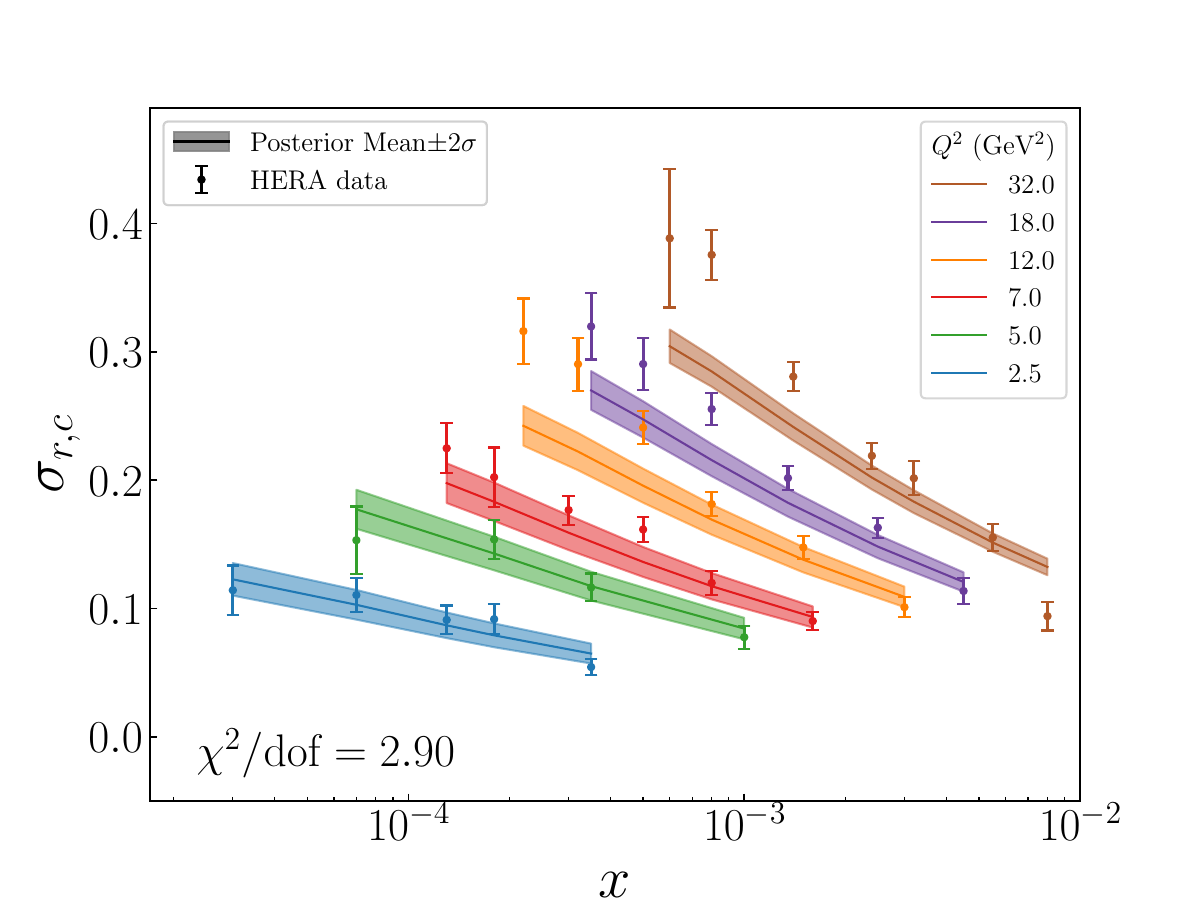}
    \caption{ Reduced total cross section and charm production as a function of $x$, in selected $Q^2$ bins at $\sqrt{s}=318 \ \gev$, with mean and $2-$sigma band calculated for posterior samples from fit, using $\balsd$ dipole running coupling prescription with an $\mathrm{MV}$ model initial condition compared to HERA data. } 
    \label{fig:modelvemulator_mv_balsd}
\end{figure*}

We begin the analysis by using the MV model parametrization for the initial condition of the BK evolution, i.e. fixing $\g=1$ in Eq.~\eqref{eq:bk-ic}, and using the Balitsky+smallest dipole running coupling prescription. The posterior distribution obtained in this case is shown in Fig.~\ref{fig:mv_balsd}.  The model parameters are tightly constrained by the HERA data.  The maximum-a-posteriori (MAP) values for the model parameters are listed in Table \ref{tab:medianmap}. The MAP values represent the values of the parameters that provide a best fit to the data (maximum value for the likelihood function). A comparison to the HERA data is shown in Fig.~\ref{fig:modelvemulator_mv_balsd}. 

We begin by noting that the MV model initial condition is insufficiently flexible to provide a good description of the HERA data. Especially the $Q^2$ dependence is faster than in the data, similarly as in leading order fits using the same functional form for the dipole amplitude~\cite{Lappi:2013zma,Casuga:2023dcf}. As $\g>1$ results in the dipole amplitude decreasing more rapidly towards small dipole sizes (probed at higher $Q^2$), it is natural to expect that having $\g$ as a free parameter would slow down the $Q^2$ evolution.

We furthermore note that the NLO fit results in relatively large proton transverse area and a small initial saturation scale. Compared to the corresponding leading order fit of Ref.~\cite{Casuga:2023dcf}, the proton transverse area is $\sim 75\%$ larger, and $Q_s^2(Y=\ln 1/0.01)=0.04\,\mathrm{GeV}^2$ which is smaller than in the leading order fit by a factor 7.
When comparing these values to the leading order analyses, it is also important to note that in Ref.~\cite{Casuga:2023dcf} the charm quark contribution was not included when computing the DIS cross section, an effect that had to be compensated by a larger  saturation scale or proton transverse area. In this analysis a relatively small value is obtained for $m_c$, resulting in a large heavy quark contribution.
We further note that the NLO result for $F_2$ is found to be roughly $50\%$ ($100\%$) larger than the corresponding structure function calculated at leading logarithmic accuracy at $Q^2=4.5\,\mathrm{GeV}^2$ ($Q^2=45\,\mathrm{GeV}^2$), i.e. the NLO correction is large (see also Refs.~\cite{Ducloue:2017ftk,Mantysaari:2022kdm} for similar estimates of the NLO correction). These results are obtained using the same  NLO BK evolved MAP parametrization for the dipole both at leading and next-to-leading order. These differences explain why a smaller $Q_s^2$ might be preferred by some NLO fits, especially if the initial condition is not flexible enough.

\subsection{MV$^\gamma$ dipole initial condition}

\begin{figure*}[ht!]
    \centering
    \includegraphics[width=0.95\linewidth]{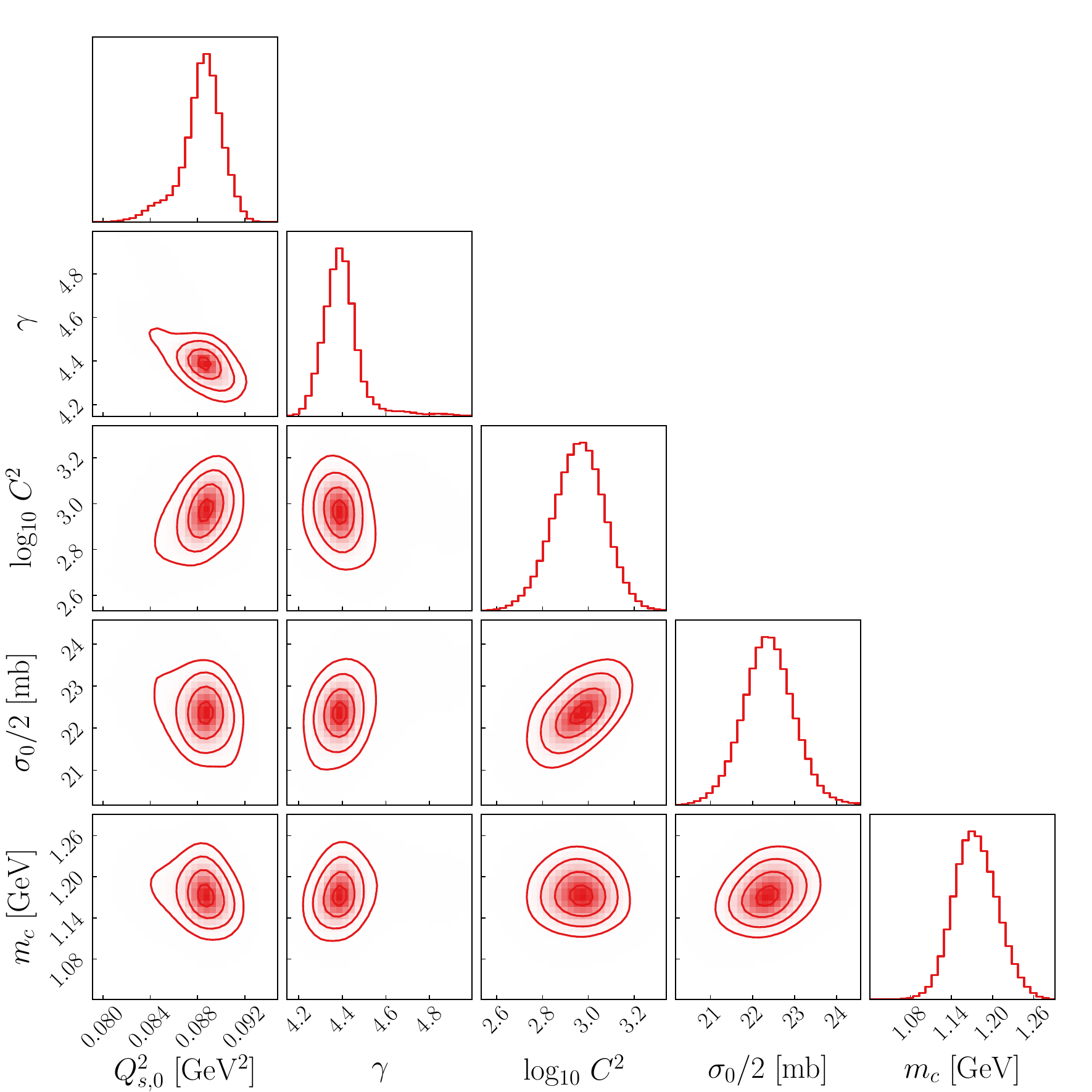}
    \caption{1- (diagonal) and 2- (off-diagonal) dimensional projections of the 5-dimensional posterior distribution found for the $\mvgamma$ initial condition using the $\balsd$ running coupling prescription.}
    \label{fig:balsd}
\end{figure*}

\begin{figure*}[ht!]
    \centering
    \includegraphics[width=0.95\linewidth]{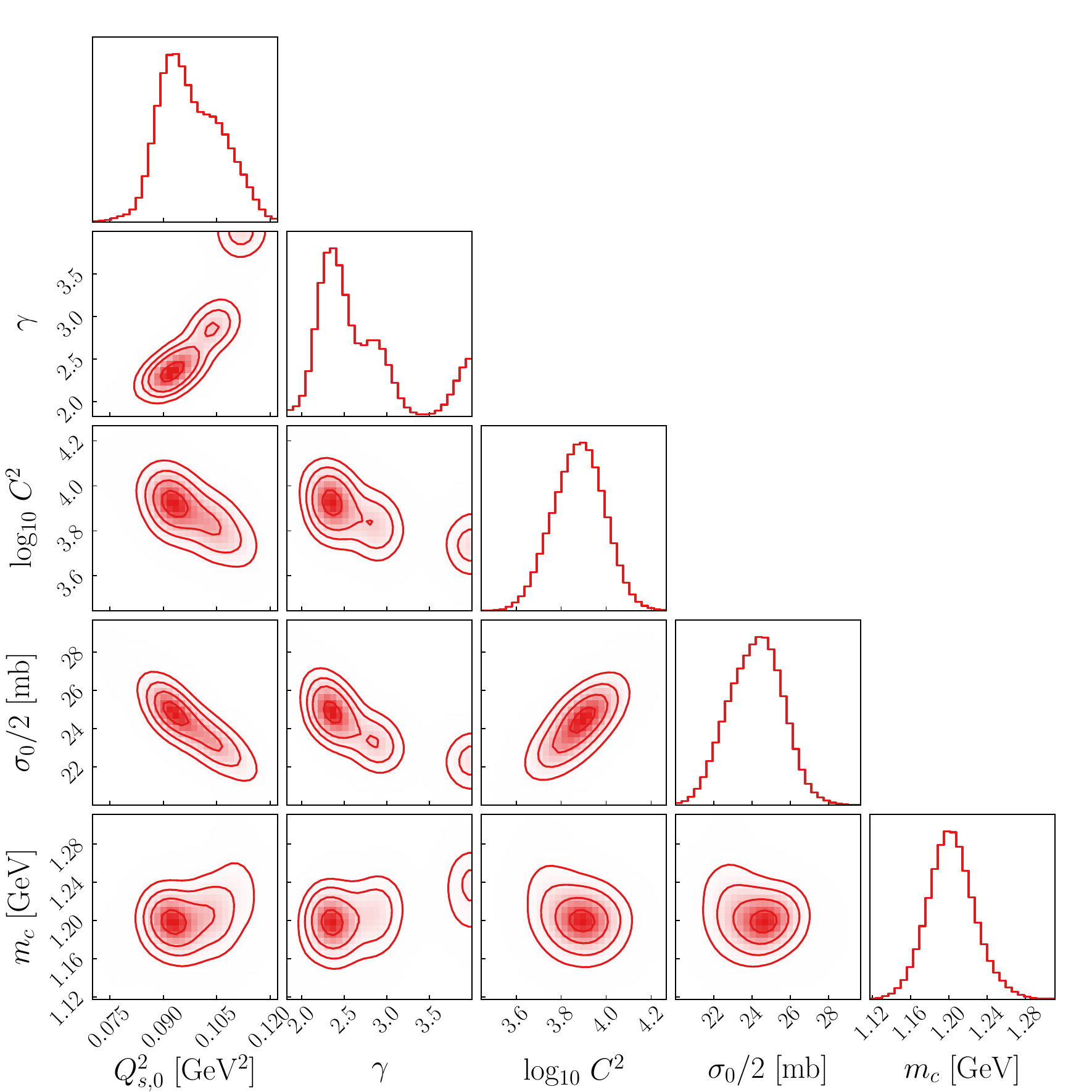}
    \caption{1- (diagonal) and 2- (off-diagonal) dimensional projections of the 5-dimensional posterior distribution found for the $\mvgamma$ initial condition using the parent dipole running coupling prescription.}
    \label{fig:pd}
\end{figure*}

The posterior distributions obtained in the Bayesian analysis with  the $\mvgamma$ initial condition ($\g$ as a free parameter in Eq.~\eqref{eq:bk-ic})  for the two different running coupling prescriptions are shown in Figs.~\ref{fig:balsd} (Balitsky+smallest dipole coupling) and~\ref{fig:pd} (parent dipole coupling). Along with that of the previously discussed MV initial condition setup,  the MAP and median parameter values for each setup are tabulated in Table~\ref{tab:medianmap}. Also indicated are the $\chisqdof$ values obtained for the MAP and median values, as well as when computing structure functions by averaging  over samples from  the posterior distribution. For all setups, we obtain median values equal to the MAP values within the numerical accuracy, which tells us that the posterior is a normal distribution and symmetric such that the 50th percentile aligns with the mode (i.e. MAP). 

Based on the $\chisqdof$ values presented, we obtain a good simultaneous description of both the constraining HERA datasets. However, similarly as in previous NLO fits~\cite{Casuga:2025etc,Hanninen:2022gje} there seems to be some tension, with $\chisqdof$ for the charm production data being of the order $2$. 

\begin{figure*}[ht!]
    \centering
    \includegraphics[width=0.49\linewidth]{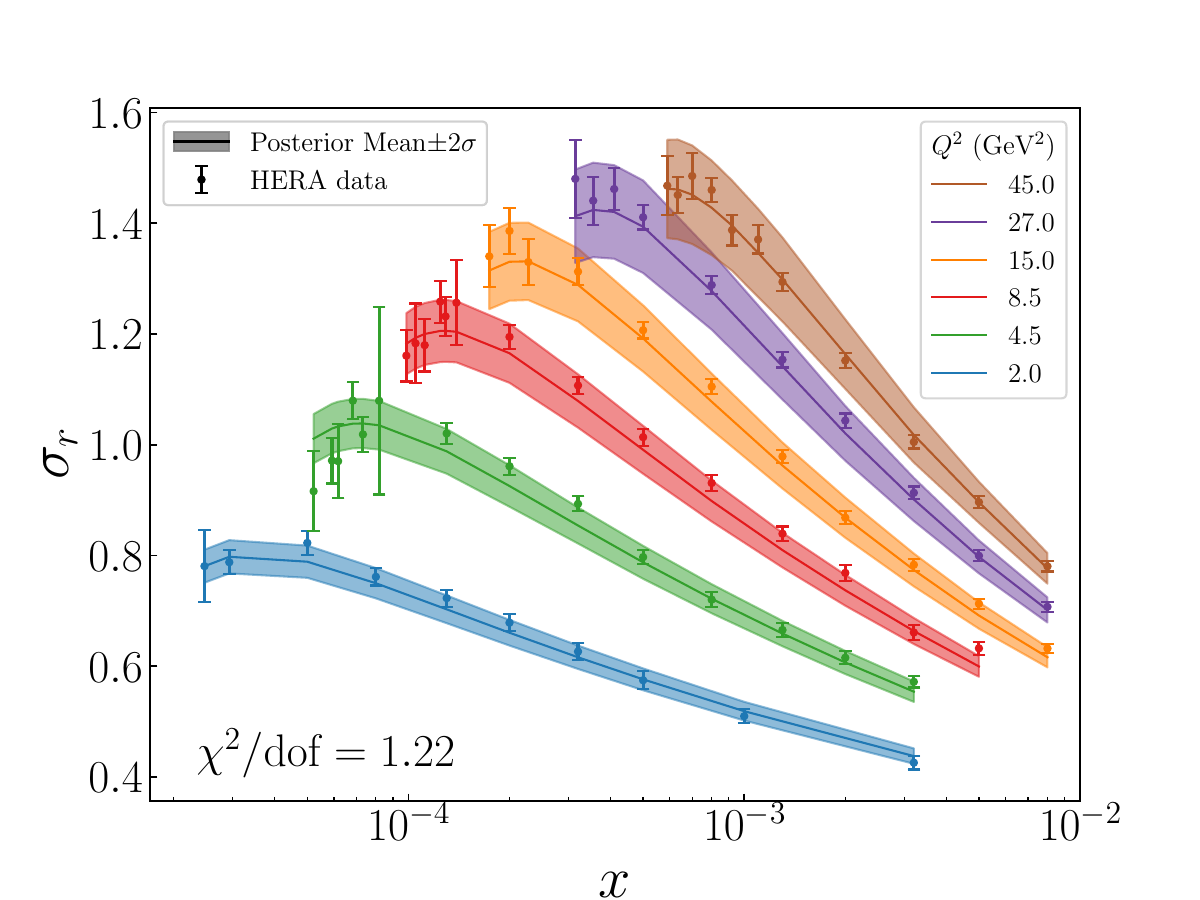}
    \includegraphics[width=0.49\linewidth]{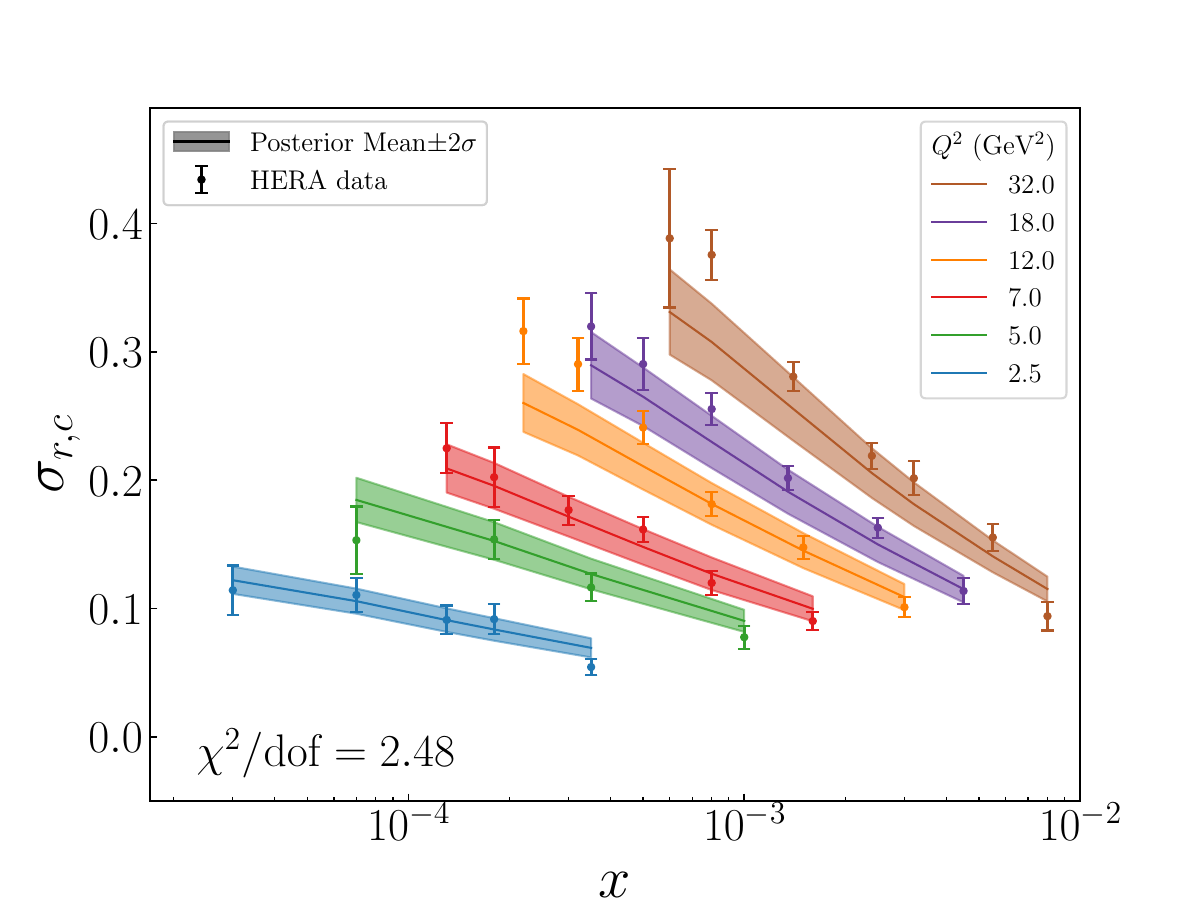}
    \caption{ Reduced total cross section and charm production as a function of $x$, in selected $Q^2$ bins at $\sqrt{s}=318 \ \gev$, with mean and $2-$sigma band calculated for posterior samples from fit, using $\balsd$ dipole running coupling prescription with an $\mvgamma$ initial condition, compared to HERA data.}
    \label{fig:modelvemulator_balsd}
\end{figure*}

\begin{figure*}[ht!]
    \centering
    \includegraphics[width=0.49\linewidth]{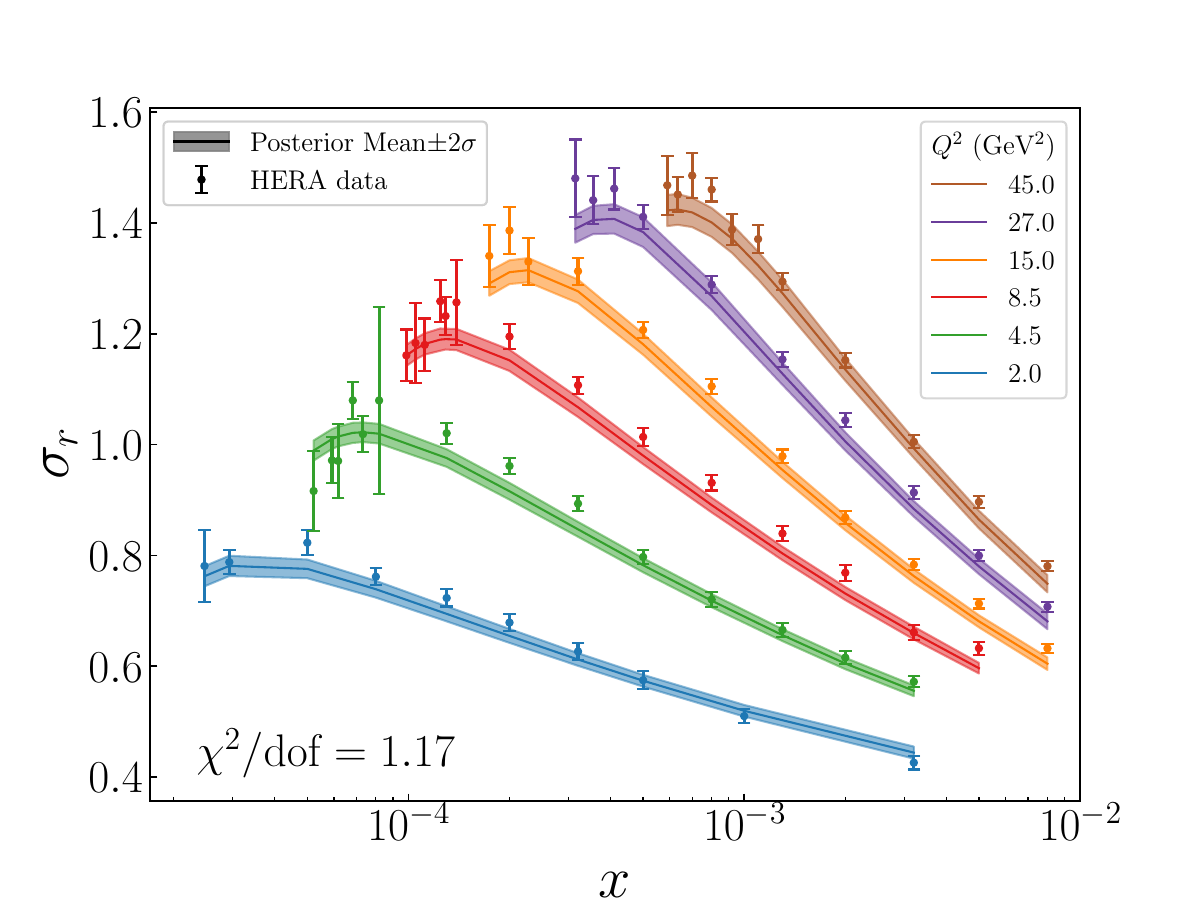}
    \includegraphics[width=0.49\linewidth]{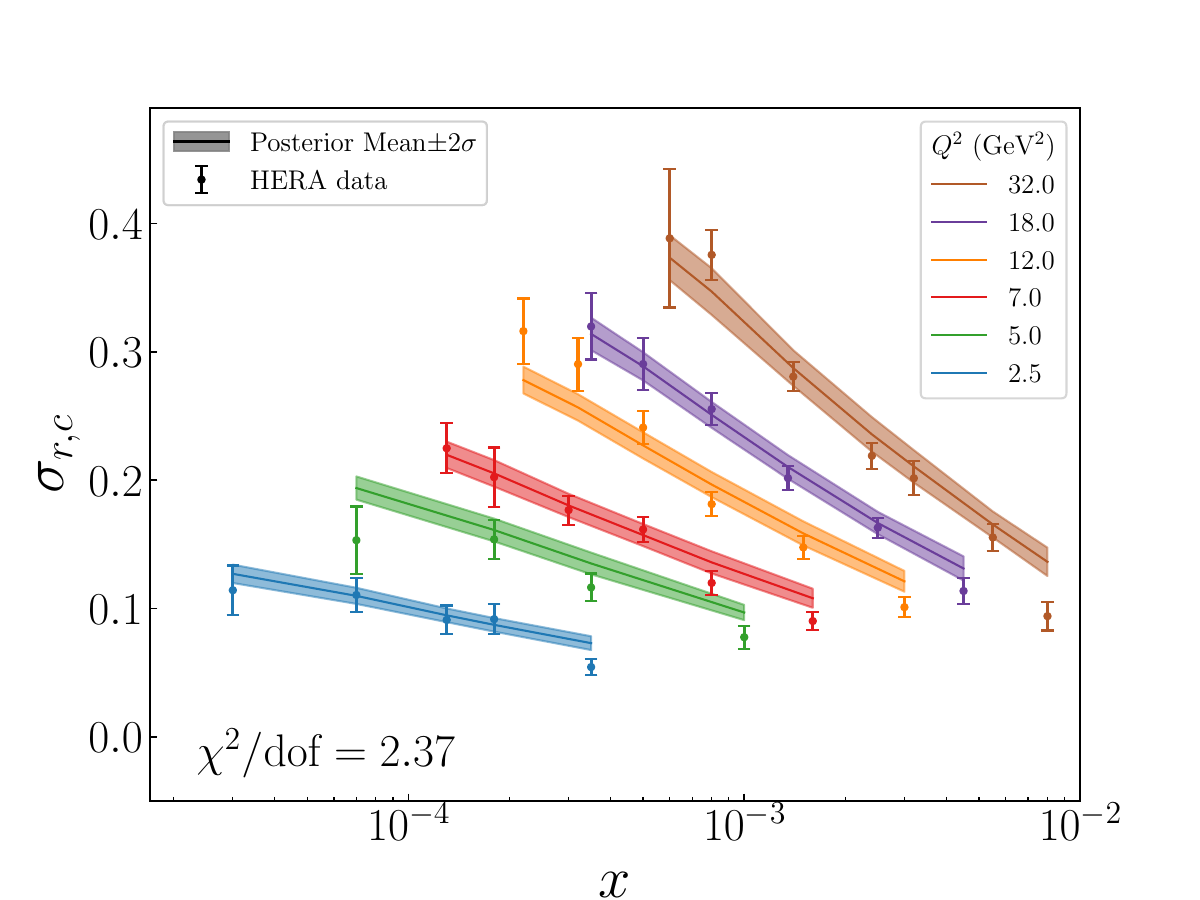}
    \caption{Reduced total cross section and charm production as a function of $x$, in selected $Q^2$ bins at $\sqrt{s}=318 \ \gev$, with mean and $2-$sigma band calculated for posterior samples from fit using parent  dipole running coupling prescription, with an $\mvgamma$ initial condition, compared to HERA data.}
    \label{fig:modelvemulator_pd}
\end{figure*}

\renewcommand{\arraystretch}{1.5}
\begin{table*}[ht]
    \centering
    \begin{tabular}{|l|c|c|c|c|c|c|}
    \hline
        \multirow{2}{*}{\textbf{Parameter Description}}  & \multicolumn{2}{c|}{\textbf{MV, Bal+SD}} & \multicolumn{2}{c|}{\textbf{$\mvgamma$, Bal+SD}} & \multicolumn{2}{c|}{\textbf{$\mvgamma$, parent dipole}}  \\ 
        
        \cline{2-7}
          & \hspace{1mm} Median $\pm \ 2\sigma$ \hspace{1mm}& MAP &\hspace{1mm} Median $\pm \ 2\sigma$ \hspace{1mm}& \hspace{1mm}MAP\hspace{1mm} &\hspace{1mm} Median $\pm \ 2\sigma$ \hspace{1mm}& \hspace{1mm}MAP\hspace{1mm} \\
    \hline
    \hline
        Initial scale, $\qso^2 \ [\mathrm{GeV}^{2}]$ &  $7.21\text{e-}3_{-0.60e-3}^{+0.62e-3}$ & $7.21\text{e-}3$ & $0.089_{-0.005}^{+0.002}$ & 0.089 &  $0.096_{-0.011}^{+0.019}$ & 0.096 \\
        
        Anomalous dimension, $\gamma$ & 1(fixed) & 1(fixed) & $4.39_{-0.11}^{+0.29}$ & 4.39 & $2.48_{-0.33}^{+1.50}$ & 2.48 \\


        Running coupling scale, $C^{2}$  & $9.83_{-2.08}^{+2.92}$ & 9.83 & $915_{-361}^{+517}$ & 915 & $7900_{-2940}^{+4240}$ & 7900 \\

        Proton transverse area, $\sigma_{0}/2 \ \mathrm{[mb]}$  & $22.8_{-2.5}^{+3.1}$ &  22.8 & $22.4_{-1.1}^{+1.1}$ & 22.4 & $24.3_{-2.5}^{+2.4}$ & 24.3 \\

        Charm mass, $m_c \ \mathrm{[GeV]}$ & $1.11_{-0.09}^{+0.09}$ & 1.11 & $1.17_{-0.05}^{+0.06}$ & 1.17 & $1.20_{-0.04}^{+0.05}$ & 1.20 \\

    \hline
    Saturation scale, $Q_s^{2}$, at $Y=\ln(1/0.01)$ $[\mathrm{GeV}^2]$ \hspace{1mm} & 0.039 & 0.039 & 0.185 & 0.185 & 0.208 & 0.208 \\ 

    \hline
    \hline
    
    \multicolumn{7}{|l|}{\textbf{$\chisqdof$ values}} \\
    \hline
        
    $\sigma_r$ data only & \multicolumn{2}{c|}{3.24} & \multicolumn{2}{c|}{1.25} & \multicolumn{2}{c|}{1.12}\\ 
    
    $\sigma_{r,c}$ data only & \multicolumn{2}{c|}{2.84} & \multicolumn{2}{c|}{2.29} & \multicolumn{2}{c|}{2.63} \\

    All data & \multicolumn{2}{c|}{3.18} & \multicolumn{2}{c|}{1.31} & \multicolumn{2}{c|}{1.21} \\
    \hline 
    \hline
    \multicolumn{7}{|l|}{\textbf{Average $\chisqdof$ over 100 samples}} \\
    \hline
    $\sigma_{r}$ data only & \multicolumn{2}{c|}{3.28} & \multicolumn{2}{c|}{1.22} & \multicolumn{2}{c|}{1.17} \\
    $\sigma_{r,c}$  data only & \multicolumn{2}{c|}{2.90} & \multicolumn{2}{c|}{2.48} & \multicolumn{2}{c|}{2.37} \\
        All data & \multicolumn{2}{c|}{3.21} & \multicolumn{2}{c|}{1.29}  & \multicolumn{2}{c|}{1.23} \\
    \hline
    \end{tabular}
    \caption{ MAP and median values, along with $2-$sigma confidence level, obtained for the $\balsd$ and the parent dipole running coupling prescriptions. Provided are model independent saturation scales at $Y=\ln \frac{1}{0.01}$ defined when $N(r^2=2/\qs^2) = 1-e^{-1/2}$. We also present $\chisqdof$ values from comparing to $\sigma_r$ and $\sigma_{r,c}$ experimental data to model calculations using the MAP and median values and the $\chisqdof$ averaged over 100 samples from the posterior.}
    \label{tab:medianmap}
\end{table*}

The HERA data constrains all model parameters well, and both running coupling prescriptions result in similarly preferred values for the model parameters $\qso^2,\initsig$ and $m_c$.
Differences in the anomalous dimension, $\gamma$, and the evolution speed parameter, $\csq$, can therefore be attributed to the difference in the running coupling prescriptions used in both setups.
The proton saturation scale $Q_s^2(Y=\ln 1/0.01)\sim 0.2$ GeV$^2$ is now also comparable to previous leading and next-to-leading order analyses\footnote{Note that we define the saturation scale as $N(r^2=2/Q_s^2)=1-e^{-1/2}$, and consequently $\qso^2 \neq Q_s^2$.}. The charm quark mass is also constrained to physically better motivated values $\sim 1.2$ GeV, compared to the MV model fit discussed above. On the other hand, the fit now prefers very large values for both the anomalous dimension $\g$ and to the running coupling scale $C^2$. The large $C^2$ indicates that the evolution needs to be slowed down significantly, given the fact that the expected value for $C^2$ is argued to be $e^{-2\gamma_E}$ in Refs.~\cite{Kovchegov:2006vj,Lappi:2012vw}. When comparing the numerical values obtained for $C^2$ one should note that in this setup the dependence on $C^2$ is only logarithmic.

Similarly to the MV model initial condition case, the proton transverse area turns out to be relatively large ($22\dots 24$ mb) compared to leading order fits where $\sigma_0/2\sim 14\dots 18$ mb~\cite{Lappi:2013zma,Casuga:2023dcf}. Previous NLO fits with the $\balsd$ running coupling prescription have also typically preferred smaller values $\sim 10$ mb.
This quantity can also in principle  be inferred from  exclusive $\mathrm{J}/\psi$ production data~\cite{H1:2013okq, ZEUS:2004yeh}, which prefers $\initsig\sim10\ldots20$ mb depending on the assumed proton density profile. The more preferable Gaussian profile represents the lower limit of this range, as well as diffractive structure function data~\cite{Lappi:2023frf}. This may point towards a possible tension between datasets if one tried to include vector meson data in a global analysis simultaneously with the structure function measurements. Furthermore, a different value for the $\initsig$ will affect the nuclear modification factor due to the saturation effects, 
as the nuclear saturation scale behaves as $Q^{2\gamma}_{s,A}(\bt) \sim \frac{\sigma_0}{2} A T_A(\bt) Q_{s,0}^{2\g}$ in the optical Glauber model~\cite{Lappi:2013zma}, see e.g. Ref.~\cite{Mantysaari:2023vfh}.

The need for large $C^2$ in our current fit can be understood as follows. The large anomalous dimension causes the dipole to, in practice, vanish quickly at $r\lesssim 1/Q_s$. As a result of the BK evolution the dipole then  evolves towards an asymptotic shape with a smooth behavior in the small-$r$ region. This causes the dipole to grow fast in the $r\lesssim 1/Q_s$ region which would result in fast growth in the DIS structure functions, compensated then by a large $C^2$ slowing down the evolution.  
Furthermore, when comparing the numerical value obtained for $C^2$ to that of the previous NLO fits~\cite{Casuga:2025etc,Hanninen:2022gje}, we note that here we are including the full NLO BK equation that results in faster evolution compared to the resummation-only approximations~\cite{Lappi:2016fmu}, and as such a larger $C^2$ is necessary in order to obtain the same $x$ dependence in HERA kinematics. Furthermore, due to the variable flavor number scheme used in $\as(\xt_{ij}^2)$,  the numerical value obtained for $C^2$ can not be directly compared to those obtained in Refs.~\cite{Casuga:2025etc,Hanninen:2022gje}.

Our results are consistent with the expectation that the $\balsd$ scheme generally results in a slower evolution since $\as(r)$ is parametrically smaller when evaluated at the scale set by the smallest dipole. This scheme will not, then, require as large $\csq$ compared to the parent  dipole scheme to obtain a good fit to the HERA data. Similar systematics was also seen in Ref.~\cite{Casuga:2025etc}. However, as mentioned above, in both cases a large $C^2$ is needed. 

Figs. \ref{fig:modelvemulator_balsd}. and \ref{fig:modelvemulator_pd} show the total reduced cross section, $\sigma_r$, and the charm quark contribution, $\sigma_{r,c}$,  calculated using samples from the posterior, where the band represents the $2\sigma$ uncertainty. The results are compared to the corresponding HERA measurements. This further shows that our results provide a good description of the available small-$x$ DIS data. 

\begin{figure*}[t]
    \centering
    \subfloat[MV, Balitsky+smallest dipole]{
    \includegraphics[width=0.33\linewidth]{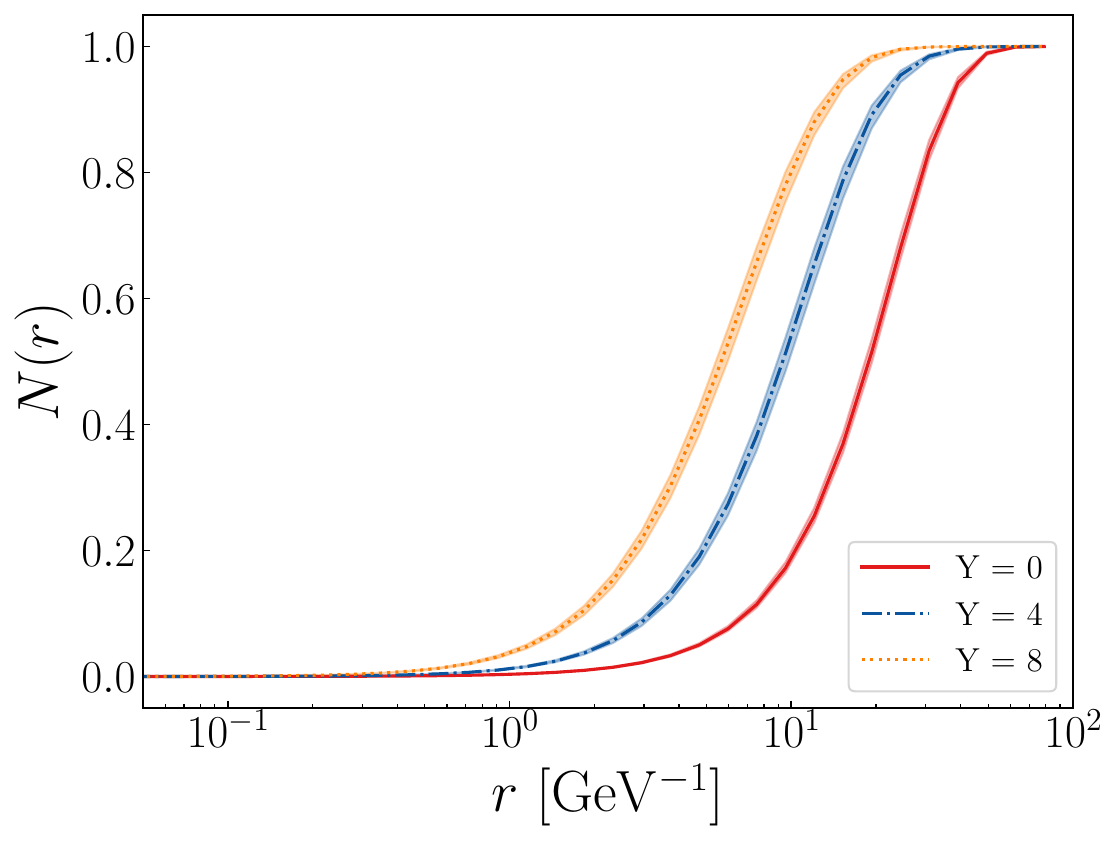}
    }
    \subfloat[$\mvgamma$, Balitsky+smallest dipole]{
    \includegraphics[width=0.33\linewidth]{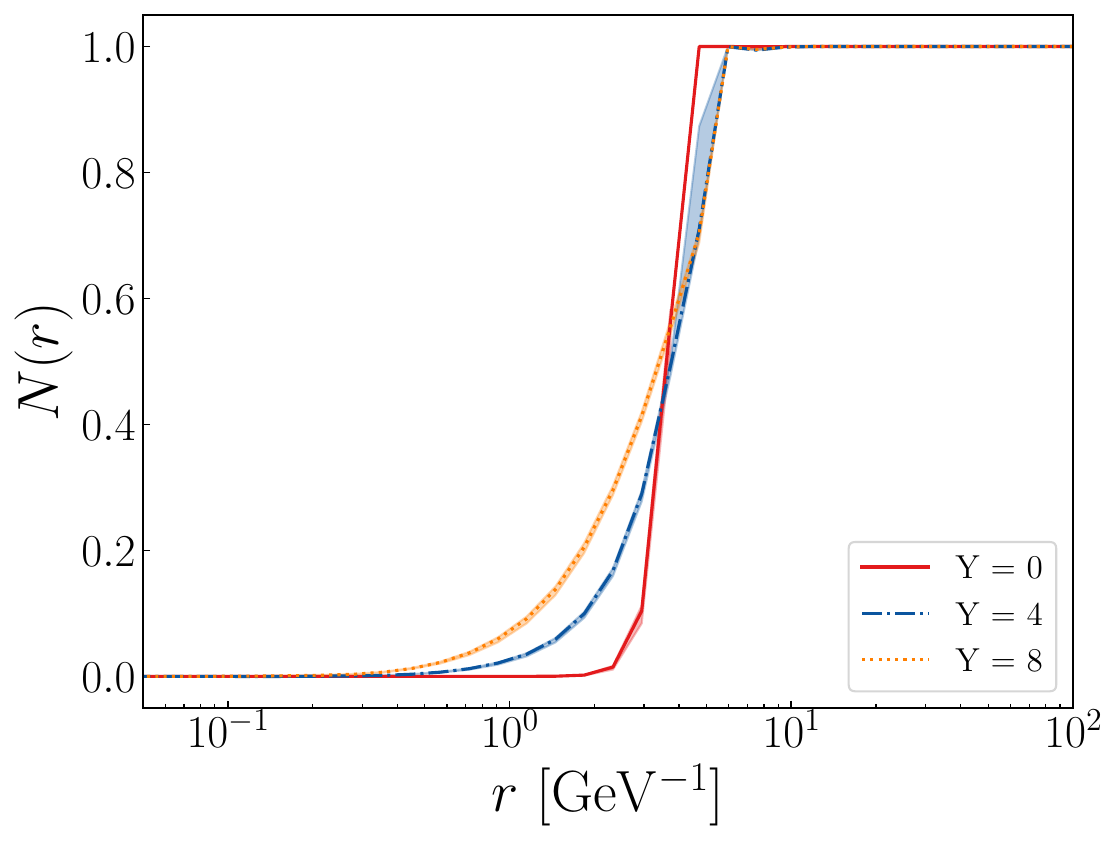}
    }
    \subfloat[$\mvgamma$, Parent dipole]{
    \includegraphics[width=0.33\linewidth]{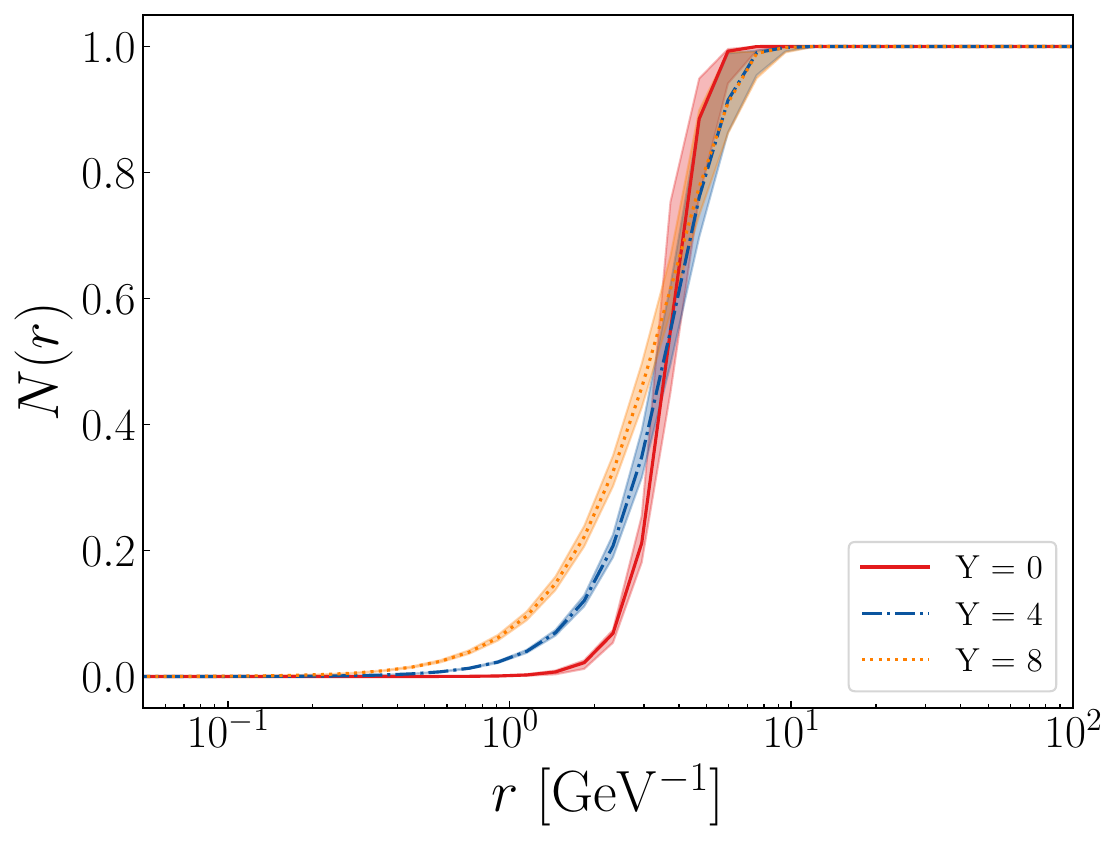}
    }
    \caption{Initial and evolved dipole amplitudes as a function of dipole size, $r = \xij{01}$, calculated from posterior samples. 
    }
    \label{fig:dipoles}
\end{figure*}

The dipole amplitude at the initial condition, and after 4 and 8 units of BK evolution, are shown in Fig.~\ref{fig:dipoles}. The effect of having a large anomalous dimension in the $\mathrm{MV}^\gamma$ initial conditions is clearly visible: at the initial $Y=0$ the amplitude is almost a step function, and a small-$r$ tail is then developed during the BK evolution. The evolution here is relatively slow due to the large $C^2$. In contrast, for the MV initial condition setup where the anomalous dimension is fixed at $\gamma=1$ and $C^2$ values are lower, the dipole amplitudes are not as steep and are evolving faster.

The initial conditions inferred for the NLO BK equation with single and double logarithmic corrections resummed to all orders are somewhat different than what was obtained in our previous NLO analysis~\cite{Casuga:2025etc} using the kinematically constrained BK (KCBK) equation. The KCBK equation is parametrically equivalent to the resummation of double transverse logarithms, but does not include single log resummation or $\as^2$ terms not enhanced by large transverse logarithms. This analysis shows that these features have a significant effect on the initial condition. This numerical importance of the single log resummation was pointed out already in~\cite{Hanninen:2021byo}.  

\subsection{Other initial conditions}

Although a good description of the HERA data is obtained using the MV$^\g$ initial condition, the resulting  values for some model parameters (especially large anomalous dimension $\g$ and slow BK evolution indicated by a large value of $C^2$) are not physically very natural. This motivates us to explore some other functional forms for the non-perturbative initial condition that could be more suitable to describe the HERA data\footnote{See also Refs.~\cite{Dai:2026nzp,Hanninen:2025iuv} for recent parameter-free extractions of the dipole amplitude.}.

In particular we have performed the same Bayesian analysis using the so called running coupling MV initial condition proposed in Ref.~\cite{Iancu:2015joa}. The rcMV model parametrization of the initial dipole amplitude is defined as 
\begin{multline}
    \label{eq:rcmv_ic}
    N_{01}( Y=0) = \Biggl\{ 1 - \exp \left[ - \left( \frac{\xij{01}^2\qso^2}{2} \bar{\alpha}_s(|\xij{01}|) \right. \right. \\
    \left.
    \left. \times \left[ 1 + \ln \frac{\bar{\alpha}_{\mathrm{sat}}}{\bar{\alpha}_s(|\xij{01}|)} \right] \right)^{p} \right]  \Biggr\}^{1/p}   
\end{multline}
where $\bar{\alpha}_s = \as\nc/\pi$ and $p$ is  additional model parameter. In this context we have also used the fastest apparent convergence (FAC) scheme for the running coupling in the BK evolution following Ref.~\cite{Iancu:2015joa}. In the FAC scheme, similarly to the Balitsky prescription, the smallest dipole parametrically sets the scale for the running coupling, and typically results in slower evolution than the Balitsky prescription. As such, one also typically expects to extract a smaller value for the $C^2$ parameter. 

However, in our setup, with the rcMV initial condition and the FAC running coupling prescription, it was not possible to obtain a good description of the HERA data ($\chisqdof \gtrsim  2.2$ for the total cross section data and $\chisqdof \gtrsim 4.0$ for the charm quark contribution), albeit preferring smaller values of $C^2\sim 10$. 

\subsection{Structure functions $F_2$ and $F_L$}
\label{sec:predictions}
\begin{figure}[h]
    \centering
    \includegraphics[width=\linewidth]{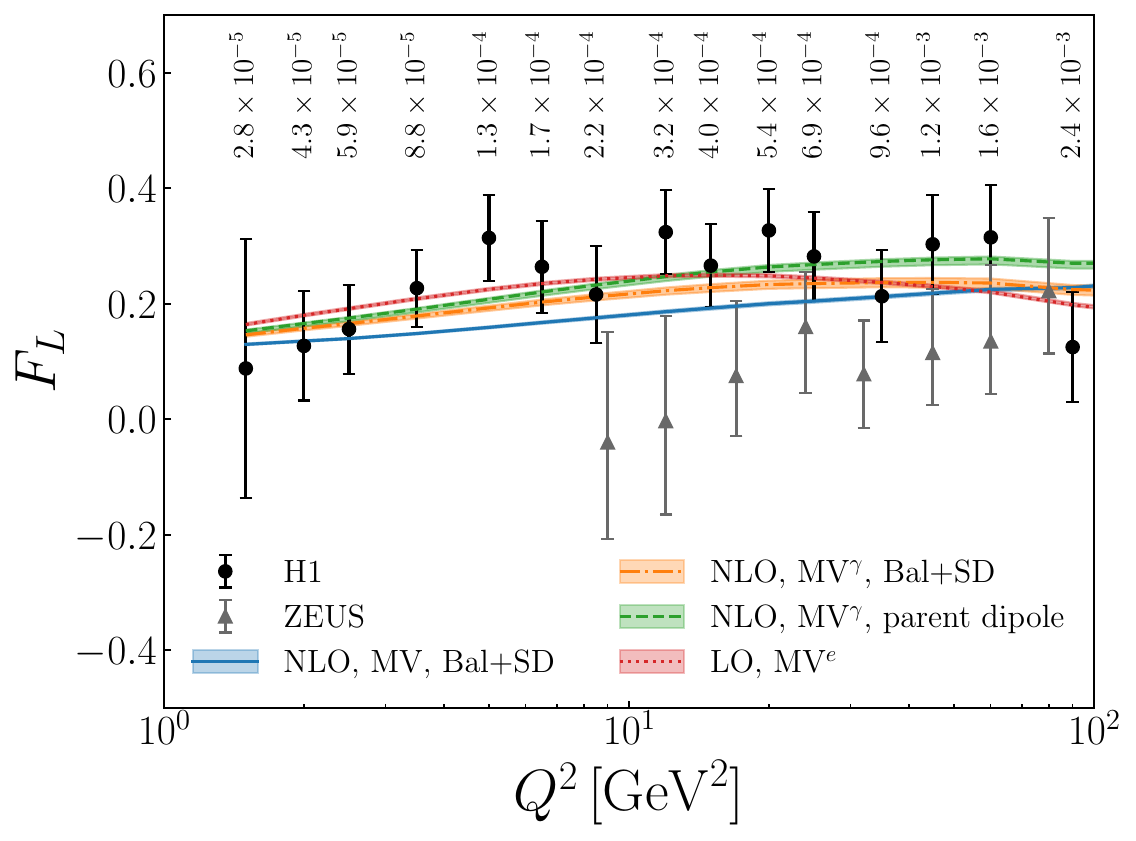}
    \caption{Longitudinal structure function calculated using different fits, compared to the HERA data~\cite{H1:2013ktq,ZEUS:2014thn}. The  $2$-sigma uncertainty estimate is also shown. The numbers in the top row indicate the Bjorken-$x$ values in the given $Q^2$ bins in the H1 data; the corresponding values for the ZEUS data are slightly different.  Leading order results from Ref.~\cite{Casuga:2023dcf} are shown for reference.
    } 
    \label{fig:fl}
\end{figure}

\begin{figure*}[ht]
    \centering
    \includegraphics[width=0.32\linewidth]{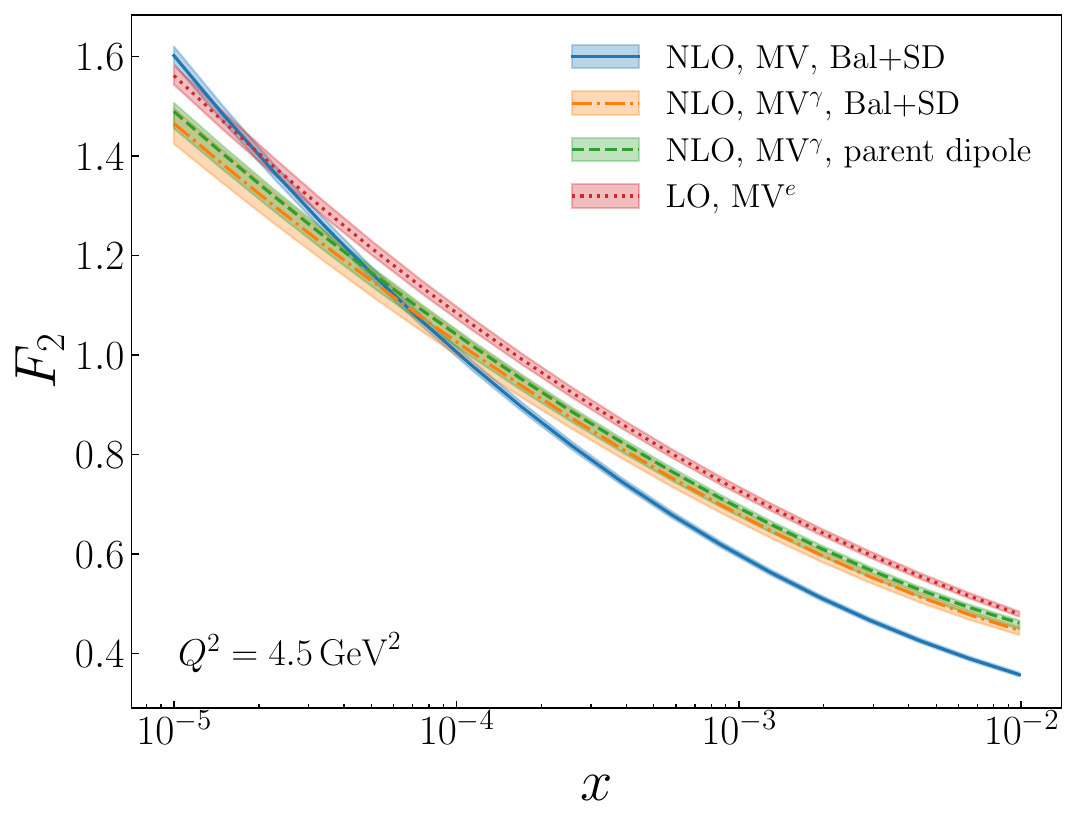}
    \includegraphics[width=0.32\linewidth]{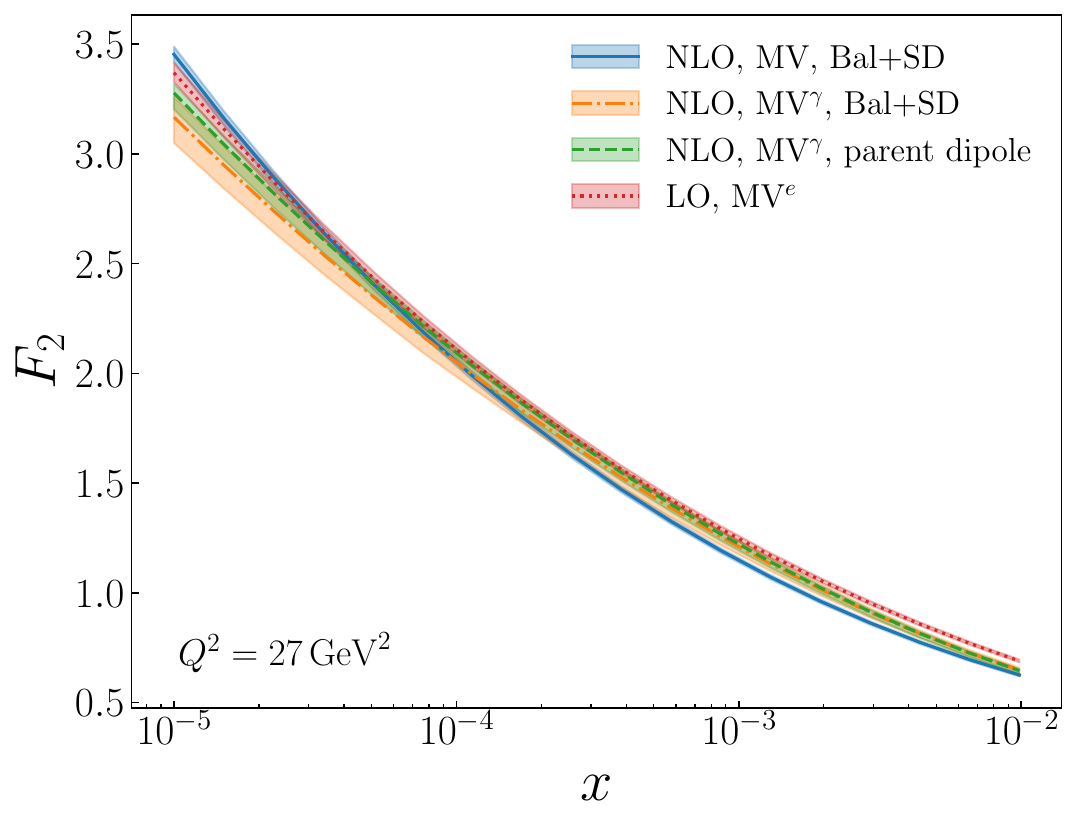}
    \includegraphics[width=0.32\linewidth]{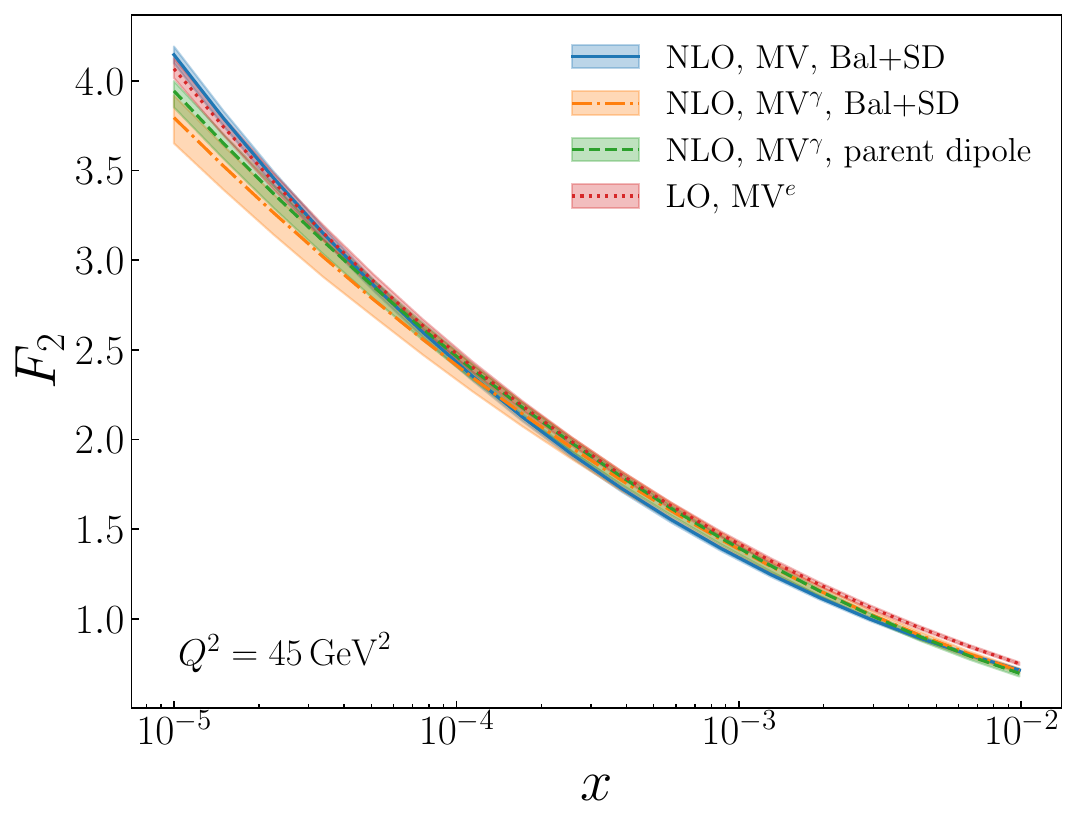}
    \caption{$F_2$ structure function as function of $x$ at different $Q^2$ calculated using different fits and averaged over posterior samples. The  $2$-sigma uncertainty estimate is also shown. Leading order results from Ref.~\cite{Casuga:2023dcf} are shown for reference.}
    \label{fig:f2}
\end{figure*}

With the inferred initial condition for the NLO BK evolution, we can compute predictions for any  observable that depends on the dipole amplitude. As an illustration, we evaluate the proton longitudinal structure function $F_L$ in HERA kinematics at $Q^2<100\,\mathrm{GeV}^2$, using the three different fits obtained in this work. The results, compared with the H1 and ZEUS data~\cite{H1:2013ktq,ZEUS:2014thn},\footnote{Note that the HERA data are binned such that  the Bjorken-$x$ increases with  $Q^2$.} are shown in Fig.~\ref{fig:fl}. For comparison, the corresponding leading order result (with only light quarks), obtained using the $\mathrm{MV}^e$ fit from Ref.~\cite{Casuga:2023dcf}, is also shown. 

All three NLO fits exhibit a similar $Q^2$ dependence to that observed in the H1 data and in the LO fit, and they provide a good description of the H1 measurements. Although some differences between the fits are visible in HERA kinematics, these remain smaller than the experimental uncertainties. Future precision measurements of $F_L$ at the EIC may nevertheless provide additional complementary constraints for global analyses.

Finally, we also present the $x$-dependence of the structure function $F_2$ in Fig.~\ref{fig:f2}. Since the reduced cross section is well approximated by $F_2$,  no significant differences between the different fits are expected in the kinematical domain covered by HERA data. Indeed, the results obtained using the NLO fits and the leading order calculation with the  MV$^e$ fit are very similar. 

At low $Q^2$, the NLO fit with the MV model initial condition exhibits a faster $x$-dependence than the other fits. This is because the $C^2$ parameter, controlling the running coupling scale, is much larger in the other two NLO fits. At higher $Q^2$, the evolution speeds are more similar. Although $C^2$ is large, the contribution from small dipoles, which is numerically more important at higher $Q^2$, increases rapidly as the steep initial condition develops a small-$r$ tail in the evolution. This is also illustrated in Fig.~\ref{fig:dipoles}. At much smaller values of $x$, the NLO fit with the MV initial condition is expected to evolve faster than the other NLO fits due to its smaller $C^2$, since the sensitivity to the initial condition is reduced in this regime.

\section{Conclusions}
\label{sec:conclusions}

In this paper, we present the first global fit to HERA structure function data within the Color Glass Condensate framework at NLO+NLL accuracy. In addition to next-to-leading logarithmic corrections to the BK equation at $\mathcal{O}(\as^2 \ln 1/x)$, we incorporate the resummation of contributions enhanced by large single and double transverse logarithms in the evolution. The proton structure functions are then obtained by convolving the BK-evolved dipole amplitudes with next-to-leading order DIS impact factors.

We extracted posterior distributions for the parameters describing the initial condition of the NLO BK equation using different running coupling prescriptions. We obtained a good simultaneous description of both the total inclusive cross section ($\chisqdof < 1.3$) and the charm quark production datasets ($\chisqdof < 2.5$). In order to achieve this agreement with the HERA data, it was necessary to use a flexible initial condition parametrization, the MV$^\gamma$ model, where the anomalous dimension $\g$ is a free parameter.
The obtained posterior distributions, available in Ref.~\cite{samples_zenodo_nlobk}, provide a statistically rigorous framework for propagating uncertainties in the initial condition to all observables within the CGC framework at NLO. By accounting for correlated experimental systematic uncertainties and simultaneously constraining with multiple datasets, we ensure a robust estimate of the uncertainty associated with the BK initial condition.

We observed the well-known preference of HERA small-$x$ data for a steep initial dipole amplitude with a large anomalous dimension $\g$. This, in turn, leads to a large effective running coupling scale. While DIS cross section calculations do not forbid large anomalous dimensions, $\g>1$ does not yield a positive definite unintegrated gluon distribution (UGD)~\cite{Giraud:2016lgg}. Functional forms that force positive UGD were also explored in this work, but those did not result in good agreement with the HERA data.

In addition to next-to-leading order corrections included in this work, in the future it will be important to quantify the importance of finite-energy (next-to-eikonal) corrections. For example, as demonstrated in Ref.~\cite{Bertilsson:2026vtu}, imposing a kinematical limit to the final state invariant mass as sizeable effects on the structure functions, especially in the heavy quark case. This could alleviate the tension between the HERA total inclusive cross section and the heavy quark distribution datasets.
Next-to-eikonal corrections to the parton-target scattering could also be implemented following related developments of  Refs.~\cite{Altinoluk:2025ang,Agostini:2024xqs,Agostini:2025vvx}.
Furthermore, it would be crucial to properly quantify the importance of the finite target size effects in $\gamma+p$ scattering, as the finite-size effects have been found to be numerically large in Refs.~\cite{Mantysaari:2018zdd,Berger:2011ew,Mantysaari:2024zxq}. It will also be interesting to include additional observables in the global analysis, for example diffractive structure functions or exclusive vector meson production, for which impact factors at NLO accuracy are available~\cite{Beuf:2024msh,Kaushik:2025roa,Mantysaari:2022kdm}.

\begin{acknowledgments}
C.C and H.M are supported by the Research Council of Finland, the Centre of Excellence in Quark Matter, and projects 338263 and 359902, and by the European Research Council (ERC, grant agreements  No. ERC-2023-101123801 GlueSatLight and ERC-2018-ADG-835105 YoctoLHC).
C.C acknowledges the support of the Vilho, Yrjö and Kalle Väisälä Foundation.
Computing resources from CSC – IT Center for Science in Finland and the Finnish Grid and Cloud Infrastructure (persistent identifier \texttt{urn:nbn:fi:research-infras-2016072533}) were used in this work.
The content of this article does not reflect the official opinion of the European Union and responsibility for the information and views expressed therein lies entirely with the authors. 

\end{acknowledgments}

\bibliographystyle{JHEP-2modlong.bst}
\bibliography{refs}

\end{document}